\documentclass{article}
\usepackage{PRIMEarxiv}

\usepackage[utf8]{inputenc} % allow utf-8 input
\usepackage[T1]{fontenc}    % use 8-bit T1 fonts
\usepackage{hyperref}       % hyperlinks
\usepackage{url}            % simple URL typesetting
\usepackage{booktabs}       % professional-quality tables
\usepackage{amsfonts}       % blackboard math symbols
\usepackage{amssymb}        % \varnothing
\usepackage{nicefrac}       % compact symbols for 1/2, etc.
\usepackage{microtype}      % micro typography
\usepackage{fancyhdr}       % header
\usepackage{graphicx}       % graphics

\usepackage{caption}
\usepackage{subcaption}
\usepackage{hyperref}
\usepackage{float}
\usepackage{todonotes}
\usepackage{amsmath}
\usepackage[toc,page]{appendix}
\usepackage{mathtools}
\usepackage{soul}
\newtheorem{proposition}{Proposition}

\title{Incentivized Network Dynamics\\ in Digital Job Recruitment

}

\author{
  Blas Kolic\\
  uc3m-Santander Big Data Institute\\
  IMDEA Networks\\
  \texttt{blas.kolic@uc3m.es} \\ 
  *corresponding author  \\
   \And
  Manuel Cebrian\\
  Center for Automation and Robotics\\
Spanish National Research Council\\
\texttt{manuel.cebrian@csic.es}\\
  \AND
  Iñaki Ucar \\
  Department of Statistics\\
  Universidad Carlos III de Madrid \\
  uc3m-Santander Big Data Institute\\
  \And
  Rosa E. Lillo \\
  Department of Statistics\\
  Universidad Carlos III de Madrid \\
  uc3m-Santander Big Data Institute\\
}

\begin{document}
\date{\today}
\maketitle

% \todo{homogenize terminology: agents/halting (general) vs. workers/job vacancies (specific)}

\begin{abstract}
Recruiting passive candidates, i.e., individuals not actively seeking jobs but open to compelling opportunities, remains one of the hardest challenges in digital recruitment. Motivated by a real collaboration with an industry partner, we introduce the \textit{Independent Halting Cascade} (IHC) model: a simple but rich agent-based framework that couples network diffusion with the possibility of halting through job applications. Agents can either recommend vacancies to peers or apply themselves, and incentives increase the likelihood of recommendation, mobilizing otherwise passive candidates. 
The IHC bridges research on social network diffusion, coordinated task completion, and labor economics by modeling heterogeneous skills, job specificities, and network structures, including homophily. We derive analytical boundaries that characterize diffusion and failure regimes, and we show, through simulations, that the IHC reproduces the empirical chain-length distributions of Travers and Milgram, and of Dodds, with only coarse calibration. Across synthetic (ER, BA, homophilic) and real networks (SMS, e-mail, Twitter), the IHC achieves comparable or higher success rates than direct-recommendation baselines, while requiring fewer applicants. 
Our findings suggest that the IHC captures core mechanisms of coordinated task completion, offering both a theoretical contribution and a practical foundation for recruitment systems designed to reach and engage passive candidates.
\end{abstract}

% keywords can be removed

\section{Introduction}
Digital platforms have reshaped how people search for and are matched to jobs, with online recruitment systems and job-matching algorithms now central to labor markets \cite{wheeler2022linkedin}. Despite these advances, recruiting \textit{passive candidates}, i.e., individuals not actively seeking a job but willing to consider attractive offers, remains a pressing challenge \cite{villeda2019use}. According to LinkedIn, about $70\%$ of potential candidates are categorized as passive and, among them, $85\%$ are open to new opportunities if approached with a compelling offer \cite{linkedin2015}. While effective in matching job openings with suitable candidates, existing recommender systems struggle to convert passive candidates due to their disinterest in job-seeking \cite{dekay2009business}. This creates a skills gap in domains such as ICT and Data Science that are in short supply of active candidates. Our work is directly motivated by a collaboration with a recruitment company that is successfully experimenting with incentive-based referral strategies to engage passive candidates, yet lacks a theoretical framework to analyze and optimize them.

Propagation through social networks has long been studied in domains such as marketing and information diffusion \cite{newman2011structure}. A foundational framework is the \textit{Independent Cascade} (IC) model \cite{goldenberg2001talk,kempe2003maximizing}, where nodes activated at each step attempt to activate their neighbors with a given probability. IC has inspired extensive work on influence maximization and seed selection \cite{saito2008prediction,waniek2020computational}, but it only models diffusion—information can spread, but it cannot be \textit{stopped} by agents making decisions to act on it.

In this paper, we extend the IC framework by introducing the \textit{Independent Halting Cascade} (IHC) model, a simple but rich agent-based model of coordinated task completion in social networks. In the IHC, agents can either propagate information to their peers or halt the cascade by applying to a vacancy, thus coupling diffusion with action. This dynamic tension—between spreading and halting—captures the essence of recruitment chains, in which information must circulate broadly but eventually culminate in a successful hire.

The IHC framework also connects to a broader literature on labor mobility and skill heterogeneity. Network perspectives show that both firm-level labor flows and individual worker mobility are shaped by the structure of social and professional ties \cite{lopez2020network,frank2024network}, while labor economics emphasizes multidimensional skill mismatch and sorting \cite{guvenen2020multidimensional,lise2020multidimensional}. By integrating heterogeneous skill sets, job-specificities, and homophilic network structures, the IHC offers a tractable microfoundation for recruitment as a coordinated, incentivized task completion problem. The role of weak ties in accessing diverse opportunities \cite{rajkumar2022causal} further motivates the study of how network topology interacts with incentives to drive successful matches. In this sense, the IHC complements emerging data-driven agent-based models of labor dynamics \cite{pangallo2024data}, focusing on the micro-mechanisms of recommendation chains and incentive diffusion.

A major novelty of the IHC lies in the integration of economic incentives that increase the likelihood of recommendations. Inspired by the literature on recursive incentive mechanisms for coordinated task completion \cite{rutherford2013limits,cebrian2012finding,rahwan2012global,naroditskiy2012verification}, we encode incentives through a parameter $\beta$ that modulates recommendation probabilities. This generic formulation captures the effect of budgeted incentives without committing to a specific allocation mechanism, and allows us to study how stronger or weaker incentives shift the diffusion–halting dynamics. In terms of recruitment, this means rewarding agents for recommending their peers, thereby mobilizing even passive candidates who might otherwise remain disengaged \cite{wheeler2022linkedin,villeda2019use,dekay2009business}.

Our main contribution is the IHC model, which, to our knowledge, is the first diffusion–halting framework explicitly designed to capture recruitment dynamics in social networks. Moreover, we derive analytical boundaries that characterize when cascades diffuse, fail, or terminate successfully, and validate them through simulations. We evaluate the model across synthetic networks (ER, BA, homophilic), heterogeneous skill settings, and real social networks (SMS, e-mail, Twitter), showing that the IHC can reproduce empirical recruitment chain distributions from Travers and Milgram \cite{travers1977experimental} and Dodds \cite{dodds2003experimental}, and can outperform direct-recommendation baselines under realistic conditions.

The remainder of the paper is organized as follows. Section \ref{sec:methods} introduces the IHC model, formalizes incentives, incorporates heterogeneous agent skills and vacancies, discusses network topologies, and defines a direct-recommendation system. Section \ref{sec:results} presents simulation results, moving from homogeneous to heterogeneous settings, and from synthetic to empirical networks. Section \ref{sec:discussion} reflects on the broader implications of IHC for digital recruitment and coordinated task completion, and outlines future research directions. All the (Python) code and reproducible results are available in our Github repository: \url{https://github.com/blas-ko/IndependentHaltingCascadeModel}

\section{Methods}\label{sec:methods}
Our main contribution is the Independent Halting Cascade (IHC) model, a diffusion–halting agent-based model that extends the well-established Independent Cascade (IC) framework \cite{kempe2003maximizing,waniek2020computational}. In the IC model, agents exclusively propagate information across the network. By contrast, the IHC model equips agents with a dual role: they may continue propagating information or halt the diffusion process by applying for a vacancy, and with some probability, successfully complete the task. This framework captures the dynamics of digital recruitment through social recommendation chains with incentives, providing the foundation for all subsequent extensions.

This Section develops in five stages. First, we present the IHC model in detail (Section \ref{sec:ihcm}). Second, we formalize incentive mechanisms that link the natural tendency to share information with the increased propensity to do so under economic rewards (Section \ref{sec:incentives}). Third, we introduce heterogeneous populations by assigning agents distinct skill sets and mapping them to job vacancies within the IHC model (Section \ref{sec:skills_vacancies}). Fourth, we explore several network topologies of varying levels of realism, including homogeneous, scale-free, and homophilic networks (Sections \ref{sec:network_topologies}). Finally, we define a direct-recommendation system to benchmark against the IHC. This system arises as a special case of the IHC model, and emulates recruitment strategies such as those used by LinkedIn (Section \ref{sec:direct_recs}). For reference, see Table \ref{tab:notation} for general terminology and notation.

\subsection{Independent Halting Cascade (IHC) agent-based model}\label{sec:ihcm}

% FIGURE!
\begin{figure}[ht!]
    \centering
    \includegraphics[width=0.6\textwidth]{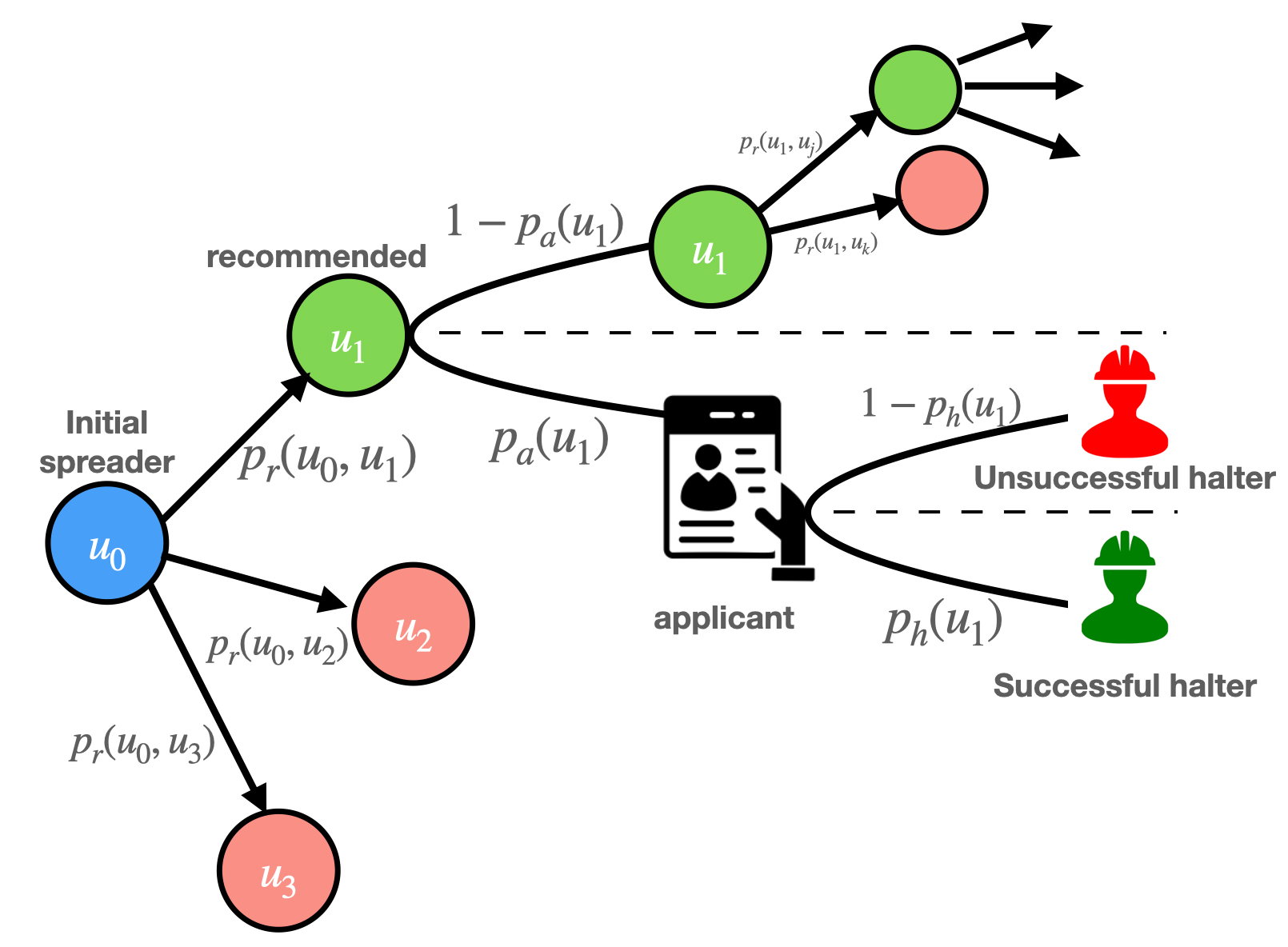}
    \caption{\textbf{Independent Halting Cascade (IHC) model diagram}: The initial spreader(s), $u_0$ (blue), recommends its neighbors, $u_1$, $u_2$, and $u_3$, with probability $p_r(u_0, u_1)$, $p_r(u_0, u_2)$, and $p_r(u_0, u_3)$, respectively. Active recommenders, here $u_1$ (green), try to recommend their neighbors with probability of $1 - p_a(u_1)$. Otherwise, they apply to halt the chain and succeed with probability $p_h(u_1)$. The cascade continues until someone successfully halts the chain (a successful chain) or there are no new recommendations (an unsuccessful chain).
    % \\
    % \textbf{Alt text:} Schematic diagram for the Independent Halting Cascade (IHC) model, where agents have the option of either propagating information or applying to halt the information cascade.
    }
    \label{fig:ihcm_diagram}
\end{figure}

The IHC model describes how information (e.g., a job posting) spreads through a social network $G = (V, E)$, where $V$ is the set of agents and $E \subset V \times V$ the potential recommendation links between them. Each agent can be in of five states $X = \{P, R^+, R^-, A, H\}$: passive ($P$), not yet involved in the process; active recommender ($R^+$), currently passing information to its neighbors; previous recommender ($R^-$), no longer recommending nor succeptible to new recommendations; applicant ($A$); attempting to halt the cascade by applying; or halter ($H$), successfully stopping the process by completing the task. For any agent $u \in V$, let $\mathcal{N}^{out}(u)$ denote its out-neighbors, and let $D_x(t)$ be the set of agents in state $x$ at time $t$. The dynamics unfold in discrete time steps as follows.

\begin{enumerate}
    \item At $t=0$, initialize the set of active recommenders as $D_{R^+}(0) = D_0$, assigning these agents to state $R^+$ (recently recommended), and set all other agents to state $P$ (passive).
\item While $D_{R^+}(t) \neq \varnothing$ and $D_H(t) = \varnothing$ (i.e., there are active recommenders and no successful halters):
    \begin{enumerate}
        \item Increment time: $t \leftarrow t+1$.
        \item Each active recommender $u \in D_{R^+}(t-1)$ activates its passive neighbors $v \in \mathcal{N}^{out}(u) \cap D_P(t)$ %; i.e., $\mathbb{P}\left( s_v(t) = R^+ | s_v(t-1) = P, s_u(t-1) = R^+, v \in \mathcal{N}^{in}(u) \right) = p^r_{uv}$. 
with probability $p_r(u,v)$. Successful activations enter $D_{R^+}(t)$, and $u$ transitions to state $R^-$. %; i.e., $\mathbb{P}\left( s_u(t) = R^- | s_u(t-1) = R^+ \right) = 1$.
        \item Each active agent $u \in D_{R^+}(t)$ applies to halt the chain with probability $p_a(u)$, transitioning to state $A$. %; i.e., $\mathbb{P}\left( s_u(t) = A | s_u(t) = R^+ \right) = p^a_u$.
        \item Each applicant $u \in D_A(t)$ halts the chain with probability $p_h(u)$, transitioning to state $H$ and successfully completing the process. %; i.e., $\mathbb{P}\left( s_u(t) = H | s_u(t) = A \right) = p^h_u$.
    \end{enumerate}
\end{enumerate}
    
In summary, the IHC process begins by initializing a set of active recommenders, while all other agents remain passive. At each step, active recommenders may try to activate their passive neighbors or attempt to apply to halt the process (e.g., applying for a job vacancy), thus transitioning out of the recommending state. If an application is made, the applicant may successfully halt the process by securing the position. The dynamics continue until no active recommenders remain (unsuccessful chain) or the chain is terminated by a successful hire (successful chain). Here, $p_r(u,v)$ denotes the probability of recommendation from $u$ to $v$, $p_a(u)$ the probability that $u$ applies, and $p_h(u)$ the probability that an applicant secures the position. See Fig. \ref{fig:ihcm_diagram} for a visual representation of the model. See  Table \ref{tab:notation} for a summary of the model states, probabilities, and other quantities.

Unlike the classical Independent Cascade (IC) model \cite{kempe2003maximizing}, agents in the IHC actively attempt to halt the cascade by completing a task, producing shorter and more contained diffusion chains compared to traditional models \cite{medya2020approximate}. In particular, if $p_a(u)=0 \ \forall u$, the IHC reduces exactly to the IC model, making IC a special case of our framework. This design reflects the assumption that agents act under incentives: they either seek to complete the task (e.g., by being hired) or benefit from helping others do so (e.g., through recommendations). In the next section, we formalize this incentive-based framework and connect it to the IHC model's coordinated task completion dynamics.

%%%% TABLE OF NOTATION
\begin{table}[ht!]
\centering
\caption{Summary of parameters, notation, and main values used in the IHC model and simulations.}
\label{tab:notation}
\begin{tabular}{p{2.2cm} p{3cm} p{8.8cm}}
\toprule
\textbf{Symbol} & \textbf{Default values} & \textbf{Description} \\
\midrule
\multicolumn{3}{l}{\textbf{Model states}} \\
$P$   & -- & Passive agent \\
$R^+$ & -- & Actively recommending agent \\
$R^-$ & -- & Previously recommended, no longer active \\
$A$   & -- & Applicant attempting to halt the chain \\
$H$   & -- & Successful hire (halts the cascade) \\
$D_x(t)$ & -- & Set of agents in state $x$ at time $t$ \\
\midrule
\multicolumn{3}{l}{\textbf{Probabilities}} \\
$p_r(u,v)$ & -- & Probability that $u$ recommends $v$ \\
$p_a(u)$   & 0.25 & Probability that $u$ applies when recommended  \\
$p_h(u)$   & 0.1 & Probability that $u$ is hired if applying \\
\midrule
\multicolumn{3}{l}{\textbf{Incentives}} \\
$\beta$ & -- & Incentive strength \\
$f_v$ & 1 & Agent $v$ fitness \\
\midrule
\multicolumn{3}{l}{\textbf{Skills and vacancies}} \\
$\mathcal{S}$ & -- & Set of all possible skills \\
$s_u$ & -- & Skill set of agent $u$ of size $n_u$ \\
$n_\nu$ & $[4,6,8]$ & Number of requirements for vacancy $\nu$ \\
$\mu_s$ & $3$ & Average number of skills per agent (Poisson parameter) \\
$M$ & $3$ & Number of latent skill groups (correlated skills) \\
$\alpha$ & $0.5$ & Dirichlet parameter for skill group correlation \\
\midrule
\multicolumn{3}{l}{\textbf{Networks}} \\
$N$ & $2000$ & Network size (number of agents) \\
$\langle k \rangle$ & $20$ & Average degree of the network \\
$\eta$ & -- & Homophily parameter \\
$\rho$ & $0.5$ & Fraction of population reached by direct recommendation \\
\midrule
\multicolumn{3}{l}{\textbf{Diagnostics and simulations}} \\
Success rate & -- & Fraction of cascades ending with a hire \\
Chain length & -- & Number of intermediaries between spreader and hire \\
Applicants & -- & Number of agents applying in a cascade \\
Simulations & $200$ & Number of runs per parameter combination \\
\midrule
\multicolumn{3}{l}{\textbf{Boundaries}} \\
Diffusion & Proposition \ref{prop:direct_halt}, Appendix \ref{app:ihc_analytical}  & Condition separating direct vs non-direct recommendations \\ 
Failure & Proposition \ref{prop:failure}, Appendix \ref{app:ihc_analytical} & Iso-curve condition for 50\% eventual failure probability \\ 
\bottomrule
\end{tabular}
\end{table}

\subsection{Incentive-based recommendation mechanism}\label{sec:incentives}

A key feature of the IHC model is that agents act under incentives: they may either propagate information to others or apply themselves in the hope of securing a reward. We formalize this behavior by adapting ideas from recursive incentive schemes, originally proposed in the context of coordinated task completion \cite{pickard2011}. In these schemes, a fixed budget $B$ is shared across the successful recruitment chain, with the halting agent receiving the largest share and the recruiters receiving diminishing rewards. This structure motivates agents not only to complete the task but also to recommend others who might, thereby coordinating the spread of information with task completion. \cite{naroditskiy2012verification, zhao2014crowdsource}

We operationalize this incentive-based behavior by integrating it directly into the recommendation probabilities. Specifically, when agent $u$ considers recommending neighbor $v$, the probability is given by
\begin{equation}
p_r(u,v) = \frac{1 - e^{-\beta  f_v p_r(u) }}{1 - e^{-\beta}},
\label{eq:incentives_pr}
\end{equation}
where $p_r(u)$ is the \textit{baseline recommendation probability} of $u$, $f_v \in [0,1]$ is the \textit{fitness} of agent $v$ for the task, and $\beta \geq 0$ is the \textit{incentive strength}. This formulation ensures that incentives always increase the chance of recommendation, except when $p_r(u)=0$ or $f_v=0$, in which case no recommendation occurs. The parameter $\beta$ implicitly encodes the effect of the incentive mechanism and the available budget: higher budgets translate into stronger incentives and thus larger values of $\beta$, provided the mechanism effectively distributes rewards. When $\beta \to 0$, then $p_r(u,v) \to f_v p_r(u)$, so the recommendation probability reverts to its baseline and becomes independent of the incentive. However, the precise way budgets and mechanisms translate into agent participation remains an open problem \cite{abdelazeem2022effectiveness, zheng2011task}; here we capture these effects generically through the parameter $\beta$, which encodes incentive strength without committing to a specific mechanism.

For simplicity, in our simulations, we set $f_v = 1$ and focus only on the role of incentives. However, this framework also allows fitness and incentive strength to guide agents in recommending neighbors who are both well-suited and rewarding to endorse.

This mechanism links the diffusion and halting dynamics of the IHC model to coordinated task completion through incentivized recommendations. The parameter $\beta$ controls how strongly incentives amplify baseline recommendation probabilities, bridging the economic context of reward distribution with the information propagation dynamics in social networks. In the next section, we extend this formulation by explicitly modeling heterogeneous skills and vacancies, which shape fitness and further influence recruitment dynamics.

\subsection{Modeling job vacancies and agent skills}\label{sec:skills_vacancies}

Next, we incorporate the compatibility of agent skills into the IHC framework to better reflect real-world recruitment, where candidates bring diverse skills and vacancies specify requirements. This affects how agents decide to apply or be hired, as it ties the application probability $p_a(u)$ and hiring probability $p_h(u)$ to the alignment between the agent's skills and the vacancy requirements \cite{mortensen1994job, neugart2018agent}. In this setting, applying agents are interpreted as \textit{job applicants}, while halters who successfully complete the task become \textit{hires}.

We define the set of all possible skills as $\mathcal{S}$, with $|\mathcal{S}| = K$ distinct skills. The \textit{skill set of an agent} $u$, denoted $s_u$, is a subset of $\mathcal{S}$ with size $n_u = |s_u|$. Similarly, a job vacancy $\nu$ is determined by its \textit{skill requirements}, which is a subset of $\mathcal{S}$ of size $n_\nu = |\nu|$. These sets are mapped to the IHC by associating them with application and hiring probabilities\footnote{Hereafter, we refer to halting probabilities as hiring probabilities.}, following established methods in which individual attributes are linked to behavioral probabilities in social networks \cite{kempe2003maximizing,saito2008prediction}. This reflects the intuition that recruitment outcomes depend on skill matching, as is the case in digital platforms \cite{villeda2019use}.

An agent $u$ is hired with probability $p_h(u)$ if their skills fully satisfy the vacancy:
\begin{equation}
    \label{eq:hiring_probs}
    p_h(u) =
    \begin{cases}
        1 & \text{if } s_u \subseteq \nu \\
        0 & \text{otherwise.}
    \end{cases}
\end{equation}

The application probability grows with the fraction of requirements met:
\begin{equation}
    \label{eq:application_probs}
    p_a(u) = \frac{| s_u \cap \nu | }{|\nu|}.
\end{equation}
More generally, any monotonically increasing function of this fraction could be used, allowing for future extensions where the mapping is learned from data \cite{guille2012predictive,yakovleva2019predict}, but we choose these functions as a simplified proxy for skill mismatch \cite{guvenen2020multidimensional}.

To generate agent skills and vacancies in simulations, we proceed as follows. First, we draw a job vacancy $\nu$ by sampling $n_\nu$ distinct skills from $\mathcal{S}$. Vacancies with larger $n_\nu$ are more specific and harder to match. Second, we sample the number of skills $n_u$ for each agent $u$ from a Poisson distribution with mean $\mu_s$.

\textbf{Uncorrelated skill sampling.} In the simplest case, we assign skills by drawing $n_u$ distinct elements uniformly at random from $\mathcal{S}$. This produces agents with unstructured skill sets and no skill correlations.

\textbf{Correlated skill sampling.} To capture realistic patterns where individual skills co-occur in different skill dimensions \cite{lise2020multidimensional, guvenen2020multidimensional}, we use a simple generative process inspired by Latent Dirichlet Allocation (LDA) \cite{blei2003latent}. Specifically, we assume $M$ latent skill groups, each corresponding to a distribution over $\mathcal{S}$. For each agent $u$, we draw a vector of group weights $\theta_u \sim \text{Dirichlet}(\alpha)$, then sample $n_u$ skills according to $\theta_u$. This produces correlated skill sets, in which certain skills co-occur more often than chance, reflecting domains such as programming languages, technical tools, or soft skills. We visualize an example of skill correlations across a population in Appendix \ref{app:homophily_plots}, Figure \ref{fig:skill_correlations}.

Finally, we compute hiring and application probabilities according to Eqs.~(\ref{eq:hiring_probs})--(\ref{eq:application_probs}).

The heterogeneous IHC model is governed by three parameters related to skills and vacancies: $\mu_s$, the average number of skills per agent; $n_\nu$, the specificity of a vacancy; and $K$, the total number of possible skills. Two additional hyperparameters are introduced when sampling correlated skills: $M$, the number of latent skill groups, and $\alpha$, the Dirichlet prior controlling the concentration of group weights. These do not alter the IHC dynamics directly, but they shape the distribution of agent skill sets and thereby influence the likelihood that vacancies can be filled.

The interplay of $\mu_s$ and $n_\nu$ determines how many agents can potentially satisfy a vacancy. If $F(x \geq n_\nu;\mu_s)$ is the Poisson cumulative probability that an agent has $n_\nu$ or more skills, and $N$ is the total population, then $N F(x \geq n_\nu; \mu_s)$ is the expected number of candidates with the sufficient number of skills to fill the job vacancy. However, since matching requires the specific skills in $\nu$, this expectation remains an upper bound. For reference, see Table \ref{tab:notation} for terminology and notation.

\subsection{Direct recommendation systems}\label{sec:direct_recs} % The oracle

We now formalize direct recommendation systems, such as LinkedIn, within the IHC framework. In these systems, a central platform or recommender has access to a large share of the population and directly prompts them to apply for a vacancy. This corresponds to a special case of the IHC model where all recommended agents apply immediately: $p_a = 1$. Thus, any agent reached in the first round of recommendations becomes an applicant without further diffusion.

Formally, we model direct recommendation as a star-like structure in which a central recommender $u^\ast$ (the initial spreader) is connected to a significant fraction $\rho$ of the population. Each connection $v$ of these $\rho N$ connections is recommended with probability $p_r(u^\ast,v)$, and if successful, the corresponding agent applies. Hiring probabilities $p_h(u)$ remain as in the IHC model and depend on the match between agent skills and vacancy requirements (Eq. \ref{eq:hiring_probs}).

This abstraction captures the fact that platforms like LinkedIn have access to a much wider set of candidates than a typical individual node in a social network. At the same time, direct recommendation lacks the diffusion and incentive mechanisms of the IHC, which can extend recruitment beyond the initially observed population. Direct recommendation systems often struggle to engage passive workers in applying for job vacancies, particularly for roles requiring highly skilled individuals \cite{wheeler2022linkedin, allal2021intelligent}. In contrast, incentive-based systems can reach people beyond the initial observed population by navigating their social networks through recommendations.

In the direct recommendation system, the probability of a successful hire depends on three factors: the number of agents who satisfy the job requirements, the fraction of those within reach of the central recommender, and the probability that at least one of them is successfully activated. We derive a closed-form expression that combines these elements in Appendix \ref{app:direct_success}, which we use throughout the simulations as an analytic benchmark for the direct-recommendation success rate.

\subsection{Network topologies}\label{sec:network_topologies}

To study the IHC dynamics under different structural conditions, we consider three types of networks. Each captures distinct features of real-world social networks, allowing us to isolate the effects of homogeneity, degree heterogeneity, and homophily on recruitment dynamics.

\textbf{Erdős–Rényi (ER) networks.} As a baseline, we use the Erdős–Rényi (ER) model $G(N,p)$, where each pair of agents is connected independently with probability $p$. This produces homogeneous networks with binomial degree distributions and no higher-order structure, serving as a null model for random connectivity.

\textbf{Barabási–Albert (BA) networks.} We then turn to the Barabási–Albert (BA) model \cite{barabasi1999emergence}, a preferential attachment process that generates networks with heavy-tailed degree distributions. These scale-free structures capture the heterogeneity of real social networks, allowing us to examine how the degree of the initial spreader influences recruitment outcomes. The BA model grows a network from $n_0$ initial nodes by sequentially adding new nodes, each with $k$ edges, preferentially attaching to high-degree nodes.

\textbf{Homophilic networks}. Finally, we construct networks with explicit homophily, in which agents are more likely to connect if their skill sets are similar. Following Talaga and Nowak’s homophily-based attachment model \cite{talaga2020homophily}, the probability of an edge between agents $i$ and $j$ is
\begin{equation}
    p_{ij} = \frac{1}{1 + \left(\frac{d_{ij}}{b}\right)^\eta},
    \label{eq:homophily}
\end{equation}
where $d_{ij}$ is the \textit{social distance} between agents $i$ and $j$, which we measure using the Jaccard similarity between their skill sets \cite{frank2024network}, $\eta$ controls the strength of homophily, and $b$ is calibrated to match a target average degree $\langle k \rangle$. For small $\eta \to 0$, homophily has little effect and edges are placed almost at random, whereas for $\eta \to \infty$, links form only if $d_{ij} < b$. In this formulation, higher homophily $\eta$ increases the likelihood of connections between agents with overlapping skills, directly linking network structure to the skill distributions introduced in Section \ref{sec:skills_vacancies}. We visualize some homophilic networks for varying levels of homophily in Appendix \ref{app:homophily_plots}, Figure \ref{fig:homophilic_networks}.

These network topologies represent varying levels of realism of social structures. ER networks provide a random baseline with homogeneous connectivity, whereas BA networks capture the heterogeneity and hub dominance typical of real social systems. Homophilic networks, on the other hand, link structure directly to agent similarity through their skills. Additionally, homophily underscores the value of weak ties: links to diverse neighbors can bridge otherwise separated clusters and expand access to specific skill sets \cite{rajkumar2022causal}. By comparing across these cases, we can separate the effects of randomness, heterogeneous, and homophilic clustering on recruitment dynamics under the IHC framework.

\section{Results}\label{sec:results}

In this section, we present our main simulation results, examining the behavior of the IHC model across increasing levels of complexity and realism. We begin with homogeneous populations (Section \ref{sec:res_homogeneous}), then incorporate heterogeneous skills and job-specific vacancies (Section \ref{sec:res_heterogeneous}), and finally move to empirical validations and real-world networks (Section \ref{sec:res_empirical}).

We develop these three stages as follows. First, we examine homogeneous settings by exploring the parameter space of the IHC model on random and scale-free networks, identifying the conditions under which diffusion produces recruitment chains beyond direct recommendations. Moreover, we analyze the effects of (economic) incentives on the recommendation probabilities. Second, we incorporate heterogeneity by assigning agents distinct skill sets and comparing the IHC process with a direct-recommendation system, showing when each mechanism is more effective. Finally, we reconstruct classic chain-length distributions from Travers and Milgram \cite{travers1977experimental} and Dodds \cite{dodds2003experimental}, and evaluate the IHC model on real networks of varying sizes and connectivities.

For all results, we run 200 simulations for each parameter combination. Unless otherwise specified, the system is initialized with a single initial spreader on social networks of size $N = 2000$ and average degree $\langle k \rangle = 20$, representing moderately connected networks \cite{newman2011structure}. Following the literature on coordinated task completion \cite{pickard2011, travers1977experimental, dodds2003experimental}, we track three diagnostic variables of the IHC: the \textit{success rate}, i.e., the fraction of successful chains; the \textit{average chain length}, i.e., the number of intermediaries between the initial spreader and the end of the chain; and the \textit{average number of applicants}, i.e., the number of agents that apply to halt during a simulation. See Table \ref{tab:notation} for a full specification of parameters and notation.

\subsection{Homogeneous parameters}\label{sec:res_homogeneous}

As a starting point, we analyze the IHC model's behavior under homogeneous parameters, where all agents share the same recommendation, application, and hiring probabilities. Formally, we set $p_r(u,v) = p_r$, $p_a(u) = p_a$, and $p_h(u) = p_h$ for every $u \in V$ and $(u,v) \in E$ in networks with expected degree $\langle k \rangle$. 

Within this homogeneous setting, we may derive an approximation of two key boundaries (see Appendix \ref{app:ihc_analytical}):
\begin{itemize}
    \item The \textit{diffusion boundary} (Proposition \ref{prop:direct_halt}), given by $\langle k \rangle p_r p_a p_h = 1$, marks the transition between regions where halts are expected to occur immediately from the initial spreader and those where recommendation chains must propagate further. 
    \item The \textit{failure boundary} (Proposition \ref{prop:failure}), defined as the parameter combinations where cascades have a 50\% chance of eventually halting successfully versus failing, separates regions of likely successful from likely unsuccessful chains.
\end{itemize}

We refer to the region just above the success boundary and maximal chain lengths as \textit{criticality}. At criticality, cascades are marginally sustained, and small parameter changes determine whether the system produces short unsuccessful attempts, long recommendation chains, or successful hires. This transition zone is particularly relevant for recruitment dynamics, as it marks the point at which incentives and network structure exert the greatest influence on outcomes.

\subsubsection{Homogeneous networks}\label{sec:res_ER}

% FIGURE!
\begin{figure}[ht!]
    \centering
    \includegraphics[width=\textwidth]{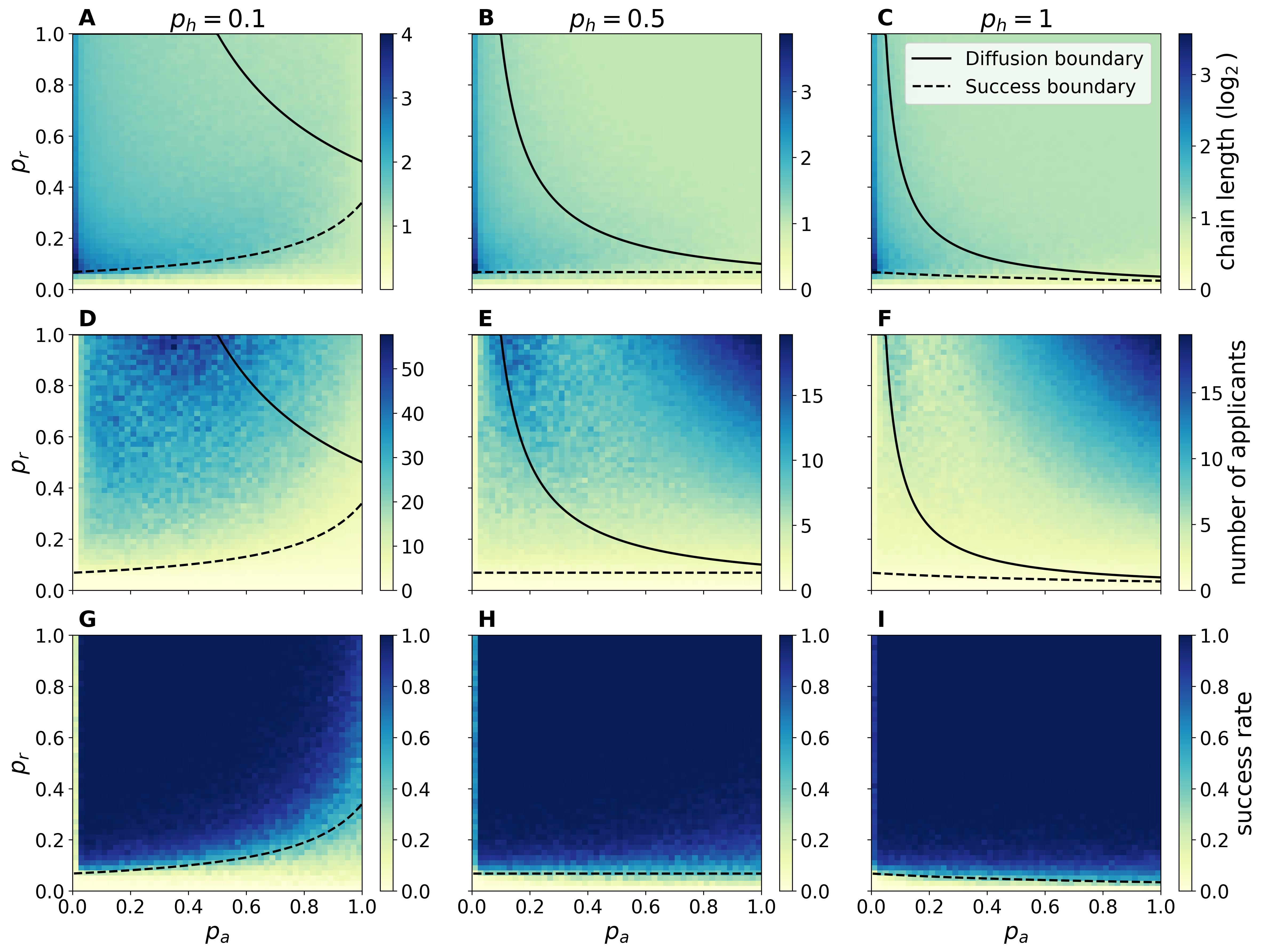}
    \caption{\textbf{IHC model behavior on Erdős–Rényi (ER) networks with homogeneous parameters.} Average chain length (top row), number of applicants (middle row), and success rate (bottom row) as functions of recommendation probability $p_r$ and application probability $p_a$, for three levels of hiring probability $p_h$ ($0.1$, $0.5$, and $1$, from left to right). The solid black line marks the \textit{diffusion boundary}, where most cascades reduce to direct-recommendation chains. The dashed black line marks the \textit{failure boundary}, below which cascades are more likely to die out without a successful halt. As $p_h$ decreases, the direct-recommendation region shrinks, and longer chains sustained by social recommendations become crucial for success.
    % \\
    % \textbf{Alt text:} Panel grid showing the behavior of the Independent Halting Cascade (IHC) model varying recommendation, application, and halting probabilities. The panels show, on the top row, the average chain length, on the middle row, the average number of applicants, and, on the bottom row, the halting success rate.
    }
    \label{fig:homogeneous_grid_ER}
\end{figure}

We begin by simulating the IHC model on Erdős–Rényi (ER) networks (see Section \ref{sec:network_topologies}) to assess its baseline behavior in the absence of social structure beyond agent connectivity. In Figure \ref{fig:homogeneous_grid_ER}, we show the IHC outcomes for a fine grid of recommendation probabilities $p_r$ and application probabilities $p_a$, selecting three representative hiring probabilities $p_h$.

Panels A–C illustrate average chain lengths. Above the diffusion boundary and below the failure boundary, most cascades reduce to \textit{direct-recommendation chains}, meaning that the initial spreader typically finds a successful halter one step away. As $p_h$ decreases (from panel C to A), the direct-recommendation region shrinks dramatically, indicating that more specific vacancies are harder to fill through direct recommendations alone. Below the diffusion boundary, chains extend beyond a single step, and chain lengths increase as diffusion through the network becomes necessary. This is the main advantage of the ICH mechanism: incentivised social recommendations become crucial to sustain the search for a successful hire.

Panels D–F show the average number of applicants, where two regions stand out. First, at high $p_r$ and $p_a$, we observe direct-recommendation dynamics in which many agents apply immediately after being contacted. The second lies around the diffusion boundary, where recommenders and applicants coexist in a delicate balance. This latter region is especially pronounced when $p_h$ is small (panel D), and it is precisely here that short but non-trivial cascades emerge. In practical terms, reaching this critical balance can lead to efficient recruitment: tasks are completed without overwhelming numbers of applicants, yet chains remain short.

Panels G–I display the success rate. Success is high across most of the parameter space once diffusion is sustained, but drops sharply below the failure boundary. As with chain lengths, low $p_h$ compresses the high-success region toward the upper-right corner of the parameter space, making direct recommendations increasingly ineffective. This demonstrates that when vacancies are highly specific, the IHC mechanism becomes essential to maintain high success rates by leveraging longer but controlled recommendation chains.

Overall, ER networks reveal the fundamental trade-offs of the IHC model: direct recommendations are enough when the recommendation and application probabilities are sufficiently high. For increasingly harder tasks, success relies on cascades sustained by social recommendations. In recruitment settings, this highlights the limitations of platform-based direct recommendation systems and the potential of incentive-driven referrals to reach candidates who are otherwise inaccessible.

\subsubsection{Effects of increasing incentives}\label{sec:res_incentives}

\begin{figure}[ht!]
    \centering
    \includegraphics[width=0.8\textwidth]{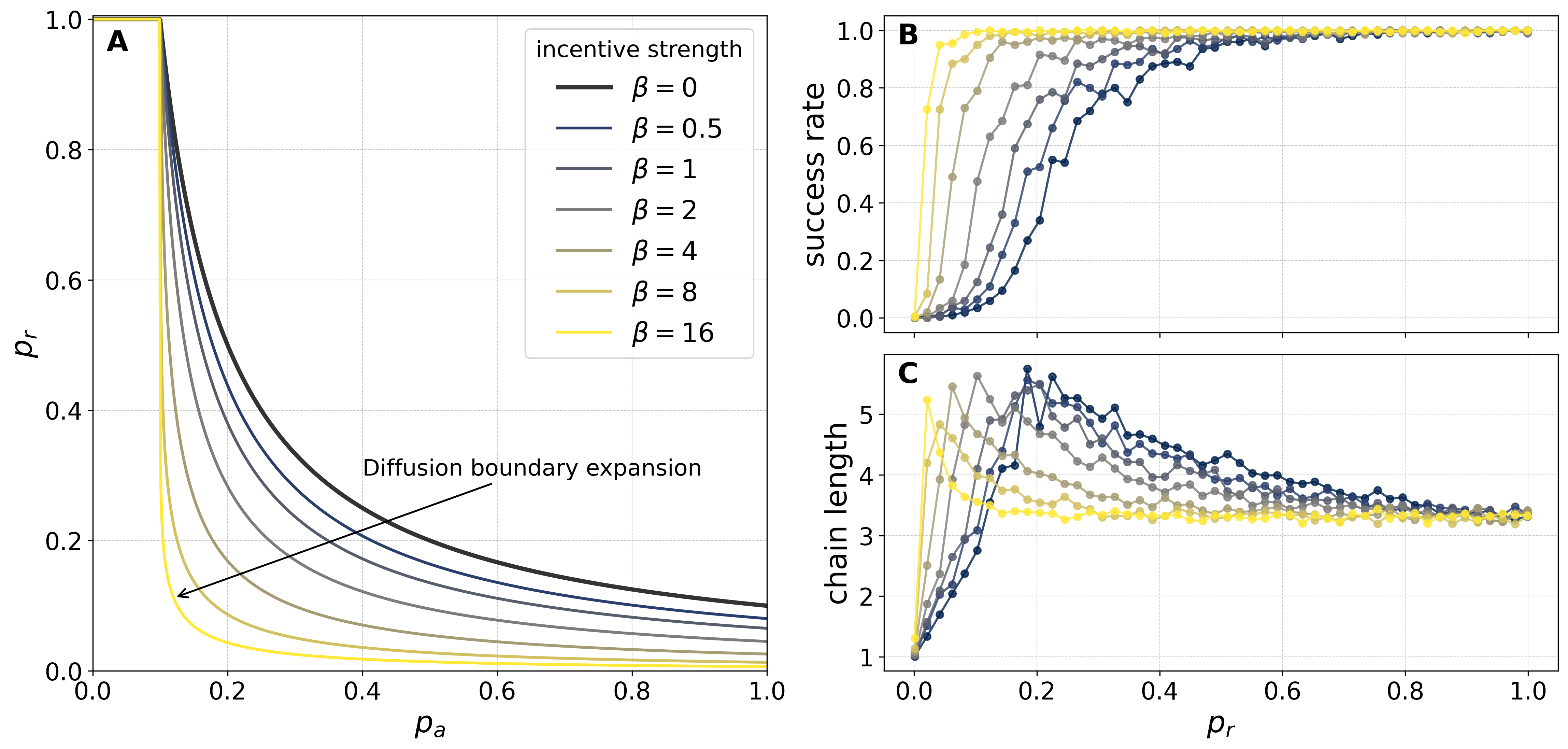}
    \caption{\textbf{Effect of increasing incentives on IHC dynamics.} \textbf{A.} Analytic transformation of the diffusion boundary under the incentive function (Eq. \ref{eq:incentives_pr}). As the incentive strength $\beta$ increases, the boundary shifts downward, reducing the baseline recommendation probability $p_r$ required for cascades to succeed. \textbf{B.} Success rate and \textbf{C.} average chain length from simulations on Erdős–Rényi networks with $N = 2000$, $\langle k \rangle = 20$, $p_h = 0.01$, and $p_a = 0.7$. Increasing incentives drive the success rate to one even at small $p_r$ values, while simultaneously shrinking the region of long cascades.
    % \\
    % \textbf{Alt text:} Plots showing the effect of increasing incentives (which increase the recommendation probabilities) in the Independent Halting Cascade (IHC) model.
    }
    \label{fig:incentives}
\end{figure}

We now study how economic incentives shape the dynamics of the IHC model. In Section \ref{sec:incentives}, we introduced Eq.~(\ref{eq:incentives_pr}), which increases recommendation probabilities through increasing (economic) incentives. This modifies the homogeneous case by replacing $p_r$ with $(1 - \exp(-\beta p_r))/(1 - \exp(-\beta))$, directly shifting both the diffusion and halting boundaries. In Figure \ref{fig:incentives}, we illustrate this effect where in Panel A we show the analytic transformation of the diffusion boundary, while in Panels B and C we present simulation results for the success rate and the average chain length in ER networks at $p_h = 0.01$ and $p_a = 0.7$, a region near the intersection of the diffusion and failure boundaries where incentives have the strongest impact.

% FIGURE!

Panel A highlights how increasing $\beta$ expands the region of direct-recommendation successful chains. As incentives grow stronger, the boundary moves downward, meaning that lower baseline recommendation probabilities are sufficient to sustain cascades. In the limit of very large $\beta$, the curve approaches a step function: once the application probability $p_a$ exceeds a critical threshold, any non-zero $p_r$ ensures direct halting.

Panels B and C highlight these effects over simulations in ER networks. As $\beta$ increases, the success rate increases dramatically, reaching one even when $p_r$ is very small (Panel B). Simultaneously, the typical transition through long chains (Panel C) shifts toward lower $p_r$, and the region of extended chains shrinks. However, for large recommendation probabilities, incentives have little influence, since chains are already short and almost certain to succeed.

These results extend the homogeneous grid analysis by showing how incentives reshape the boundaries that organize IHC dynamics. When vacancies are highly specific and direct recommendations alone are unlikely to succeed, incentives enlarge the region of successful cascades and suppress the prevalence of long chains, while leaving high-recommendation regimes largely unaffected.

\subsubsection{Effects of initial spreader on scale-free networks}\label{sec:res_BA}

Up to this point, we have employed the ER model to generate random networks, resulting in agents with highly homogeneous connectivity, making the initial spreader selection largely irrelevant. However, many real social networks exhibit heterogeneous degree distributions that follow power-law or exponential distributions \cite{barabasi1999emergence, hernando2010unravelling}, suggesting that the connectivity of initial spreaders could play a crucial role in the IHC model's ability to identify and hire a suitable agent for a given job vacancy.

This section investigates how networks with more realistic, heterogeneous degree distributions impact the IHC model's behavior. We turn to the Barabasi-Albert (BA) model \cite{barabasi1999emergence}, a preferential attachment dynamic network model that generates realistic power-law distributions and captures the self-organizing behavior observed in real social networks (see Section \ref{sec:network_topologies}). We perform simulations at $p_h = 0.1$ and $p_a = 0.25$, following Travers and Milgram's suggested dropout rate \cite{travers1977experimental}. % The BA model expands a network with $n_0$ initial nodes by adding new nodes, each with $k$ edges, preferentially attaching to existing nodes with high degrees.

% FIGURE!
\begin{figure}[ht!]
    \centering
    \includegraphics[width=\textwidth]{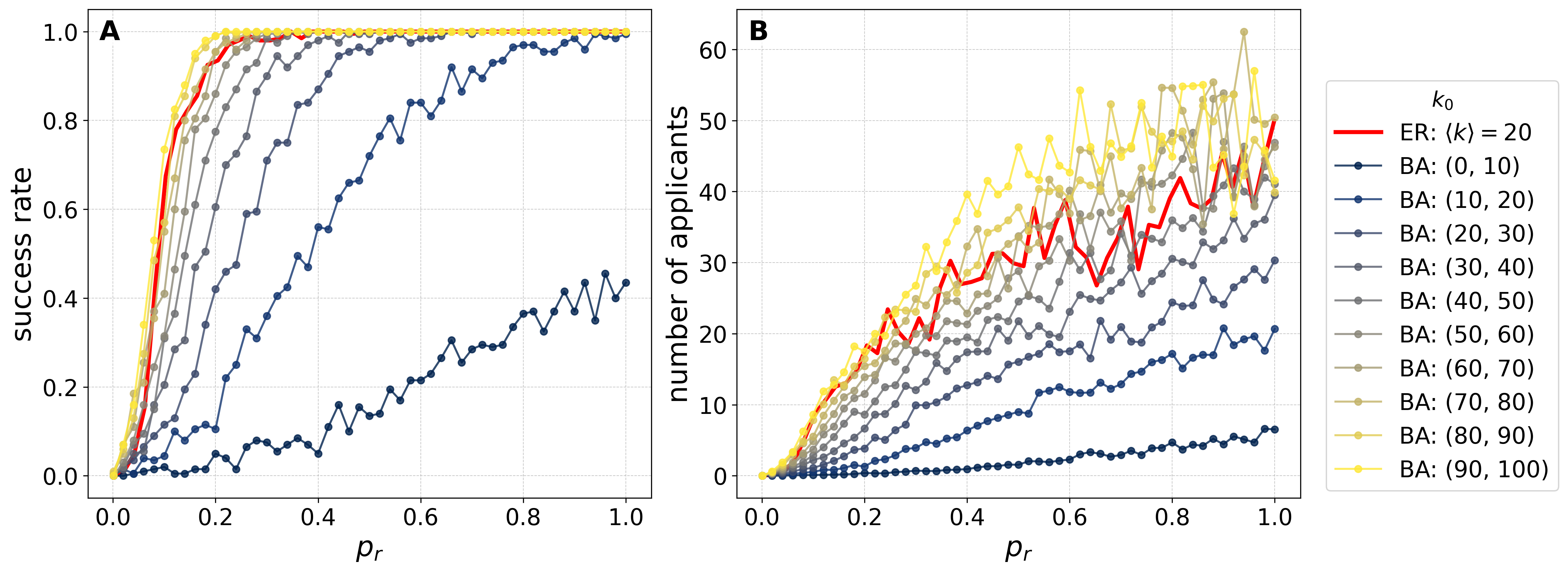}
    \caption{\textbf{Effect of the initial spreader’s connectivity on IHC dynamics in BA and ER networks.} \textbf{A.} Success rate and \textbf{B.} number of applicants as functions of the recommendation probability $p_r$ for Erdős–Rényi (ER) networks with $\langle k \rangle = 20$ (red, solid) and Barabási–Albert (BA) networks conditioned on the initial spreader’s degree $k_0$ (shaded lines), with $p_h = 0.1$ and $p_a = 0.25$. Higher $k_0$ values in BA networks increase both the success rate and the number of applicants.
    % \\
    % \textbf{Alt text:} Effect of initial speader degree on the success rate (left) and number of applicants (right) of the Independent Halting Cascade (IHC) model for Barabasi-Albert networks.
    }    
    \label{fig:heterogeneous_network_results}
\end{figure}

In Figure \ref{fig:heterogeneous_network_results}, we show how the initial spreader connectivity, $k_0$, shapes recruitment outcomes in BA networks and compare it to ER networks with $\langle k_0 \rangle = 20$. In Panel A, success rates increase with both $k_0$ and the recommendation probability $p_r$. When the initial spreader is poorly connected ($k_0 \in [0, 10]$), the success rate remains low and does not converge even when $p_r=1$. As $k_0$ increases, the success rate increases steadily, eventually reaching $100\%$ in all cases, converging faster for larger $k_0$. In contrast, ER networks achieve near-perfect success even in small $p_r$, highlighting that homogeneity in degree distributions makes successful cascades more likely even when the average connectivity is the same.

From Panel B, we observe that the total number of applicants increases roughly linearly with $k_0$, suggesting that in BA networks, the degree of the initial spreader largely dictates overall IHC outcomes. BA and ER networks produce comparable numbers of applicants when the spreader connectivity is similar, but poorly connected spreaders in BA networks generate far fewer applicants and often fail to trigger successful cascades. Overall, these results indicate that the initial spreader’s connectivity is a key determinant of IHC success, independent of other network characteristics.

\subsection{Heterogeneous agent skills and job vacancies}\label{sec:res_heterogeneous}

We now extend the analysis to heterogeneous populations, where agents possess distinct skill sets and vacancies have specific requirements. This setting captures more realistic recruitment scenarios, as application and hiring depend not only on the recommendation process but also on skill–vacancy compatibility (see Section \ref{sec:skills_vacancies}).

Unless noted otherwise, heterogeneous simulations use an average number of skills per agent $\mu_s = 3$, and vacancies with $n_\nu \in \{4,6,8\}$ requirements. These correspond to progressively more specific jobs, which are satisfied by roughly 50\%, 10\%, and 1\% of the population, respectively. When sampling correlated skills, we use $M=3$ latent skill groups with a Dirichlet prior with parameter $\alpha=0.5$. See Table \ref{tab:notation} for a summary of the parameters and notation.

We organize the heterogeneous IHC results into two parts. First, we analyze \textit{homophilic networks}, where edges are more likely between agents with similar skill sets (see Section \ref{sec:network_topologies}). Then, we compare the \textit{heterogeneous IHC} to a \textit{direct-recommendation} system under the same skill and vacancy specifications (see Section \ref{sec:res_ihc_vs_direct}), identifying parameter regimes where each mechanism is more effective.

\subsubsection{IHC vs direct recommendation system}\label{sec:res_ihc_vs_direct}

We compare the heterogeneous IHC model of social incentives with a direct-recommendation system, in which the initial spreader directly reaches a fixed fraction of the population. Specifically, the direct system assumes access to $\rho = 0.5$ of the agents, with the reachable set chosen at random in each simulation. For the IHC model, we use ER networks with an average of $\langle k \rangle = 20$ contacts, and skills are sampled without correlation.

% FIGURE!
\begin{figure}[ht!]
    \centering
    \includegraphics[width=\textwidth]{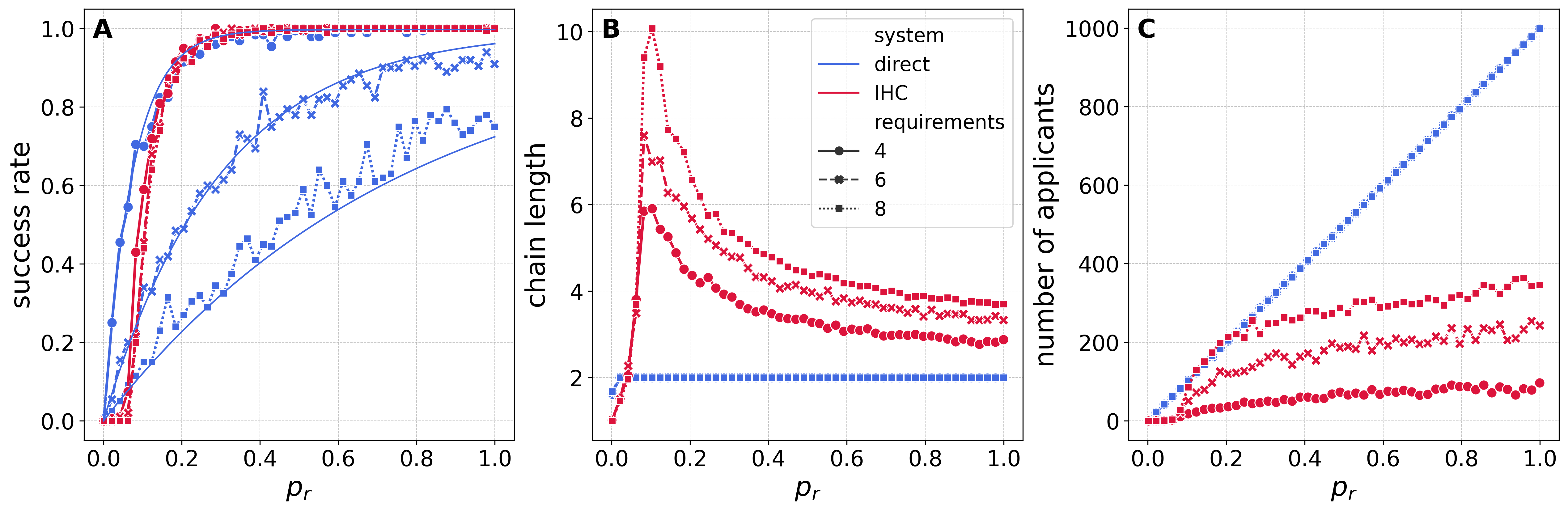}
    \caption{\textbf{Comparison of IHC (red) and direct-recommendation systems (blue) in ER networks.}  \textbf{A.} Success rate, \textbf{B.} average chain length, and \textbf{C.} average number of applicants as functions of the recommendation probability $p_r$ for vacancies with $n_\nu \in \{4,6,8\}$ requirements. The direct recommendation system has access to half of the population ($\rho = 0.5$), always produces chains of length two, and yields a linear increase in applicants with $p_r$. The IHC achieves comparable success rates even for specific vacancies, at the cost of longer but more efficient chains with fewer applicants.
    % \\
    % \textbf{Alt text:} Comparison of the Independent Halting Cascade (IHC) model against direct recommendation systems on Erdos-Renyi networks for different outcomes: success rate (left), average chain length )middle) and average number of applicants (right) for a population with heterogeneous skill sets and increasingly specific job vacancies.
    }
    \label{fig:ihc_vs_direct_ER}
\end{figure}

In Figure \ref{fig:ihc_vs_direct_ER}, we show the success rate (panel A), average chain length (panel B), and number of applicants (panel C) for the IHC model (red) and the direct-recommendation system (blue) as a function of the recommendation probability $p_r$ for increasingly specific job vacancies.

In panel A, we observe that both systems improve their success rates as $p_r$ increases. However, they exhibit contrasting behaviors regarding the vacancy specificity, $n_\nu$. For the direct system, higher specificity substantially reduces the success rate. Respectively for $n_\nu = \{4,6,8\}$, the maximum success at $p_r=1$ drops to approximately $\{0.97, 0.65, 0.31\}$, reflecting that success is limited by whether the reachable half of the population contains a suitable candidate. Moreover, we plot the analytic success rate of the direct-recommendation system based on Eq. (\ref{eq:direct_analytical_solution}) in Appendix \ref{app:direct_success}, which closely aligns with simulations. By contrast, the success rate in the IHC model converges to one, regardless of $n_\nu$, provided $p_r$ is sufficiently large. This highlights the role of peer-to-peer recommendations in overcoming the sparsity of highly specific skill matches.

Panel B explains this difference by showing that more specific vacancies require longer recommendation chains in the IHC model. While the direct system always produces chains of length two (spreader plus applicant), the IHC generates longer cascades when vacancies are harder to fill. Once $p_r$ exceeds a threshold (around $0.1 - 0.2$), most chains become successful, and their average length decreases as $p_r$ grows. This mechanism allows the IHC to maintain high success even for rare skill combinations.

Finally, in panel C, we show the average number of applicants. As expected, this number grows linearly with $p_r$ in the direct system, quickly reaching large values because many recommended agents apply simultaneously. In the IHC model, growth is much slower because only a subset of agents apply along the recommendation chains. This makes the IHC more efficient, leveraging the social network rather than the brute-force approach of the direct system, which attempts to reach half the population at once.

\subsubsection{Effect of homophilic agent connections}\label{sec:res_homophilic}

We now examine the IHC process in \textit{homophilic networks}, where links are more likely to form between agents with similar skills (see Section \ref{sec:network_topologies}). To generate these networks, we sample correlated skills from $M=3$ latent groups with Dirichlet parameter $\alpha=0.5$ and set the job specificity to $n_\nu = 6$, meaning that only about $10\%$ of the population possesses the required number of skills. Thus, skills are heterogeneous and highly correlated across groups. We generate networks using the homophily-based attachment method of Talaga and Nowak \cite{talaga2020homophily}, with the homophily parameter $\eta$ ranging from $\eta=0$ (corresponding to ER networks) to $\eta=\infty$ (a hard threshold model where edges exist only between highly similar agents). This setup enables us to isolate how network structure interacts with correlated skills in the IHC model. We visualize some of these networks in Appendix \ref{app:homophily_plots}, Figure \ref{fig:homophilic_networks}.

% FIGURE!
\begin{figure}[ht!]
    \centering
    \includegraphics[width=0.8\textwidth]{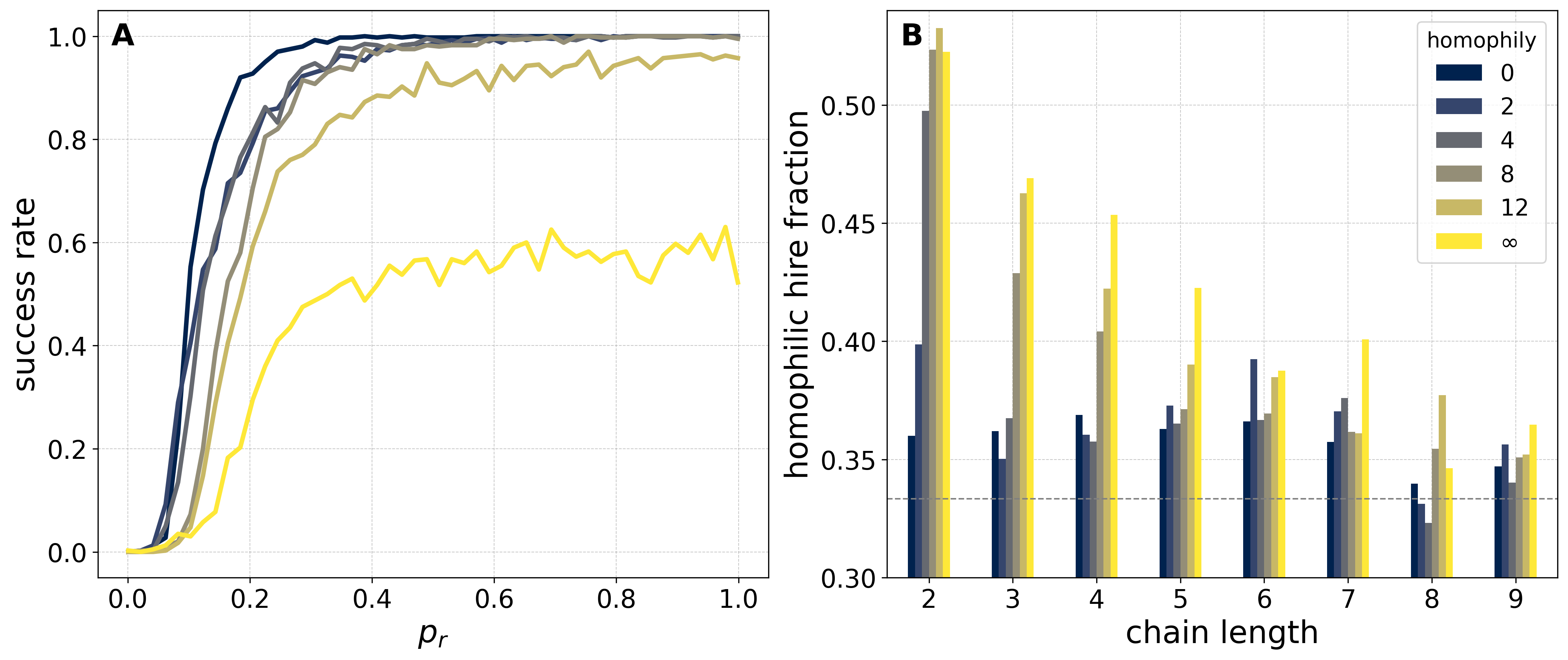}
    \caption{\textbf{IHC outcomes in homophilic networks with correlated skills.} \textbf{A.} Success rate as a function of the recommendation probability $p_r$ for increasing homophily levels $\eta$ ($\eta = 0$ represents ER networks), with $n_\nu = 6$. Success remains high under low or moderate homophily but declines sharply for very high homophily. \textbf{B.} Fraction of hires belonging to the same latent skill group as the initial spreader, conditioned on successful chain length. Homophilic networks show a strong within-group bias for short chains, but this fraction converges to the $1/M$ baseline for longer chains, reflecting the role of weak ties.
    % \\
    % \textbf{Alt text:} Outcomes of the Independent Halting Cascade (IHC) model for homophilic networks, including (left) halting success rate, and (right) the proportion of homophilic hires as a function of the chain length.
    }
    \label{fig:homophilic}
\end{figure}

In panel A of Figure \ref{fig:homophilic}, we show the success rate as a function of the recommendation probability $p_r$ for several homophily levels. For low to moderate homophily, the curves are nearly indistinguishable from those of random networks, indicating that success is largely unaffected by homophilic connections. However, at very high homophily ($\eta=12$ and especially $\eta=\infty$), success rates drop sharply, even at $p_r=1$, suggesting that successful hires are hard when the network is too clustered around agents with similar skills, limiting exposure to diverse skill sets.

To further explore this, we measure whether successful hires belong to the same skill group as the initial spreader. In panel B of Figure \ref{fig:homophilic}, we report the fraction of such \textit{homophilic hires}, conditioned on successful chains at different lengths. In ER networks, this fraction stays slightly above the baseline $1/M$, reflecting random exposure across groups. In contrast, homophilic networks show a strong bias toward within-group hires at short chain lengths. However, as chains grow longer, the fraction of homophilic hires steadily declines and converges toward the baseline. This illustrates the role of weak ties \cite{rajkumar2022causal}, where longer cascades bridge otherwise distant parts of the network, enabling access to candidates outside the initial spreader’s skill group.

Overall, these results highlight a key trade-off. Random networks (or networks with weak homophily) facilitate rapid exploration of diverse candidates, sustaining high success rates. Strongly homophilic networks, by contrast, trap diffusion within skill-similar clusters, reducing success unless cascades are long enough to escape. In the context of job recruitment, weak ties are crucial for connecting with candidates with complementary skills, particularly when vacancies are highly specialized.

\subsection{Empirical analysis}\label{sec:res_empirical}

For our last set of results, we simulate the IHC model against empirical benchmarks and real-world social networks. This serves to validate its ability to reproduce observed recruitment dynamics and to explore how heterogeneous skills and vacancies interact with the structure of actual networks.

First, we revisit the classic experiments of Travers and Milgram \cite{travers1977experimental} and Dodds et al. \cite{dodds2003experimental}, showing that even in its homogeneous form, the IHC can replicate the empirical distributions of successful chain lengths under mild calibration. We then extend the analysis to heterogeneous agents embedded in real social networks of increasing size and connectivity, highlighting how vacancy specificity, skill distributions, and network structure jointly shape the outcomes of the IHC compared to the direct-recommendation system.

\subsubsection{Reproducing empirical chain length distributions}\label{sec:res_chain_lengths}

We test whether the IHC model reproduces the chain-length distributions observed in classic recruitment experiments. We focus on Travers and Milgram’s small-world study \cite{travers1977experimental} and the later large-scale replication by Dodds et al. \cite{dodds2003experimental}.

To calibrate the model, we fix $p_h = 0.1$ and $p_a = 0.25$, the dropout rate reported in \cite{travers1977experimental}. Using the bisection method, we vary $p_r$ in ER networks until the simulated average chain length of successful chains matches the empirical average. This gives $\hat{p_r} = 0.128$ for Milgram (simulated: $5.11$, empirical: $5.16$) and $\hat{p_r} = 0.18$ for Dodds (simulated: $4.14$, empirical: $4.15$). The corresponding success rate places the Milgram case closer to the critical intersection of the failure and diffusion boundaries, at 75\%. Meanwhile, the Dodds case is further away from criticality, with consistently shorter chains and a success rate of 93\%.

% FIGURE!
\begin{figure}[ht!]
    \centering
    \includegraphics[width=0.8\textwidth]{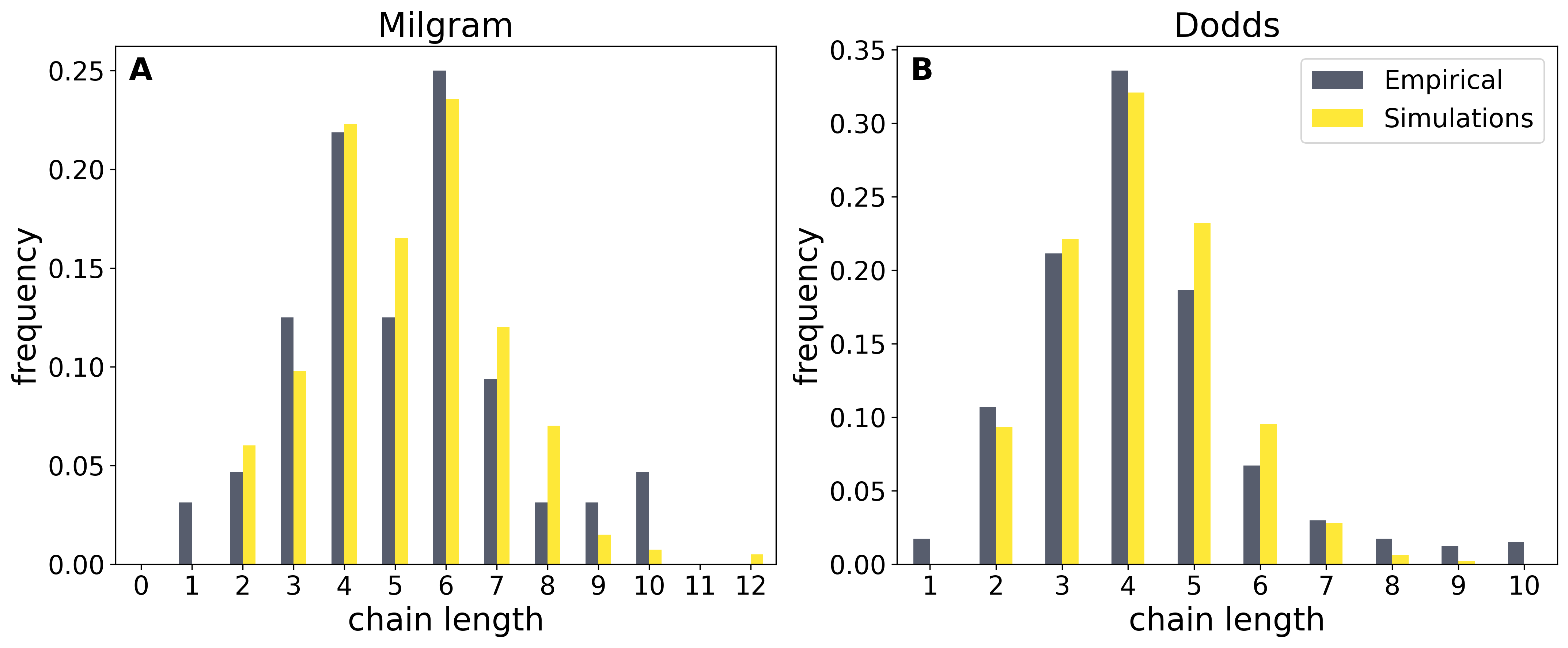}
    \caption{\textbf{Reproducing empirical chain length distributions.}  Simulated IHC distributions (blue) calibrated to match the average chain length of successful cascades in the classic experiments of \textbf{A.} Travers and Milgram \cite{travers1977experimental} ($\hat{p_r}=0.128$) and \textbf{B.} Dodds et al. \cite{dodds2003experimental} ($\hat{p_r}=0.18$). Despite calibrating only the mean, the simulated distributions closely follow the empirical ones (gray).
    % \\
    % \textbf{Alt text:} Reproduction of the empirical chain length distributions using the Independent Halting Cascade (IHC) model for (left) the Travers and Milgram 1977 experiment, and (right) Dodds et al 2003 experiment.
    }
    \label{fig:empirical_chains}
\end{figure}

In Figure \ref{fig:empirical_chains}, we compare the empirical and simulated chain-length distributions at $\hat{p_r}$. Remarkably, the IHC reproduces the overall shape of the distributions despite being calibrated only to match the average chain length. This suggests that the IHC model captures the core mechanisms driving real-world recruitment dynamics, even under simple homogeneous assumptions.

\subsubsection{IHC simulations on real networks}\label{sec:res_real_nets}

% FIGURE!
\begin{figure}[ht!]
    \centering
    \includegraphics[width=\textwidth]{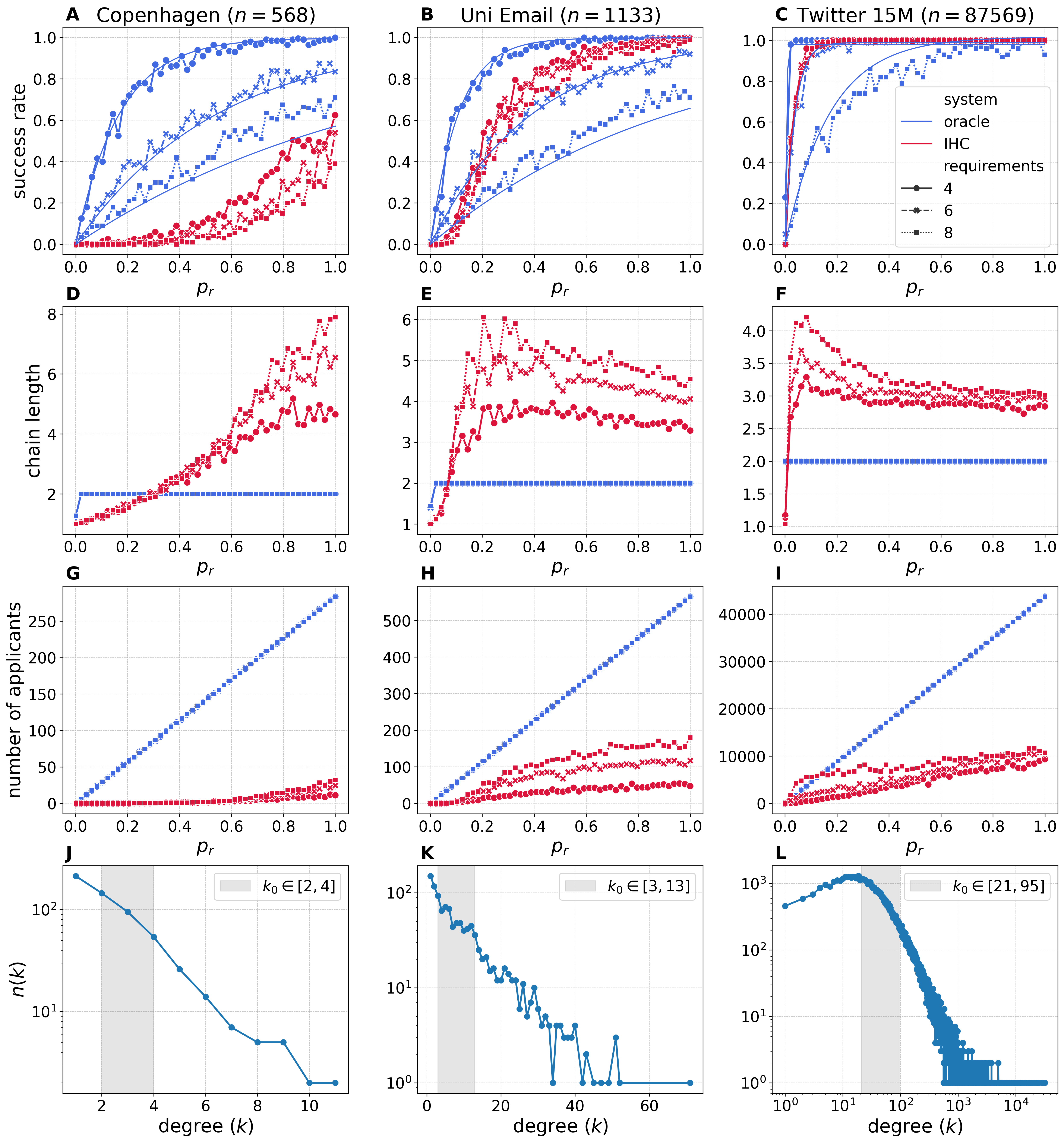}
    \caption{\textbf{IHC and direct recommendation on real networks.}  \textbf{top.} Success rate, \textbf{upper middle.} average chain length, \textbf{lower middle.} average number of applicants, and \textbf{bottom.} degree distribution as a function of $p_r$ for three empirical networks: Copenhagen SMS (left), University Email (center), and Twitter (right). Sufficiently Well-connected networks favor the IHC, which achieves high success with far fewer applicants, independent of the job specificity $n_\nu$, while in sparse networks, direct recommendations perform better.
    \textbf{Alt text:} Comparison of the Independent Halting Cascade (IHC) model against direct recommendation systems on several empirical complex networks for different outcomes: success rate (left), average chain length (middle), and average number of applicants (right) for a population with heterogeneous skill sets and increasingly specific job vacancies.}    
    \label{fig:ihc_vs_direct_real}
\end{figure}

We now test the IHC model on three empirical networks of increasing size and connectivity: the Copenhagen SMS network ($N=568$, $\langle k \rangle = 2.4$) \cite{stopczynski2014measuring}, the University Email network ($N=1133$, $\langle k \rangle = 9.6$) \cite{guimera2003self}, and the Twitter protest network ($N=87{,}569$, $\langle k \rangle = 107.5$) \cite{gonzalez2011dynamics}. In each case, we compare the heterogeneous IHC against a direct-recommendation system with access to half of the population ($\rho=0.5$), using uncorrelated skills with $\mu_s=3$ and vacancy specificities $n_\nu \in \{4,6,8\}$ (see Table \ref{tab:notation}).

We present our results in Figure \ref{fig:ihc_vs_direct_real}, which shows, for each network and as a function of $p_r$, the success rate (top), the average chain length (upper middle), the average number of applicants (lower middle), and the empirical degree distribution (bottom) for the IHC and the direct-recommendation systems. Additionally, we plot the analytic success rate of the direct-recommendation system based on Eq. (\ref{eq:direct_analytical_solution}) in Appendix \ref{app:direct_success}, which closely matches the simulation results. In each case, we sample initial spreaders by selecting their degree $k_0$ from the $25-75$ quantile range of their degree distribution (shaded region in the bottom panels).

The Copenhagen network, with very sparse connectivity, is well below the critical boundary, where success rates remain low for the IHC across all $p_r$. Direct recommendations consistently dominate, highlighting that in poorly connected networks, diffusion chains cannot reliably reach qualified candidates.

The University Email network lies closer to criticality. The IHC achieves high success rates as $p_r$ increases, but convergence is slower and requires longer chains, particularly for very specific job vacancies ($n_\nu=6,8$). Direct recommendations perform better only for the least specific vacancies, with a much higher number of applicants than the IHC in all cases.

Finally, the Twitter network is far above criticality, so both systems quickly achieve near-perfect success rates, each with distinct patterns. Direct recommendations are more effective for low-specificity vacancies, whereas the IHC's performance is independent of vacancy specificity. In all scenarios, the IHC requires far fewer applicants, underscoring the efficiency of incentive-driven diffusion in large, highly connected networks.

These findings demonstrate that network connectivity is crucial in shaping the relative performance of centralized versus decentralized recruitment. While the direct-recommendation system excels in smaller, less connected networks (the Copenhagen SMS network), the IHC system becomes increasingly effective in medium-sized (Uni Email) and large networks (X/Twitter). In these larger networks, the decentralized nature of the social system allows it to outperform direct recommendations, especially as the recommendation probability increases. Each network's degree distribution plays a key role in the recruitment efficiency, with centralization becoming less advantageous as network size and agent connectivity increase.

\section{Discussion}\label{sec:discussion}
In this study, we introduced the Independent Halting Cascade (IHC) model, a framework that integrates network diffusion with coordinated task completion through economic incentives to address the challenge of recruiting passive job candidates. The IHC extends the well-established Independent Cascade model \cite{kempe2003maximizing} by allowing agents not only to propagate information but also to halt its spread by applying for a vacancy. In doing so, the model captures the trade-off between sharing opportunities with peers and acting on them individually. This halting mechanism, combined with recursive incentives inspired by coordinated task-completion challenges \cite{pickard2011}, provides a novel approach to modeling recruitment and similar tasks in which both information flow and individual action are crucial.

Our analyses showed that the IHC generates a variety of behavior-rich regimes that go well beyond traditional diffusion dynamics. By exploring the space of recommendation, application, and hiring probabilities, we identified distinct boundaries that determine whether cascades grow, halt directly, or fail altogether. These regimes proved useful in understanding recruitment scenarios: when hiring probabilities are low (corresponding to highly specific vacancies), the region of direct recommendations shrinks, and network diffusion driven by incentives becomes essential. Conversely, when vacancies are less specific, direct recommendations suffice, but at the cost of engaging more applicants. This ability to tune and compare regimes highlights the flexibility of the IHC framework.

A key strength of the model lies in its adaptability across different network topologies. On Erdős–Rényi networks, the IHC recovered classical percolation-like transitions, while on scale-free Barabási–Albert networks, the degree of the initial spreader proved decisive in determining success rates and cascade lengths. In homophilic networks constructed from correlated skills, we showed that high homophily reduces success rates but also revealed the importance of weak ties in bridging across groups, in line with empirical findings \cite{rajkumar2022causal}. These results resonate with recent evidence that mobility and recruitment are constrained by network structure \cite{frank2024network}. Finally, by comparing the IHC with a direct recommendation system, we demonstrated that while direct recommendations are efficient for broad vacancies or in sparse networks, the IHC consistently outperforms them for specific vacancies and in larger, more connected networks.

Notably, the IHC was able to reproduce empirical chain length distributions from Travers and Milgram \cite{travers1977experimental} and Dodds \cite{dodds2003experimental}. By calibrating only the average chain length of successful cascades, the model reproduced not just the mean but also the full distribution with remarkable fidelity. This suggests that the IHC captures some of the underlying mechanisms of coordinated task completion through incentivized recommendations, even under homogeneous assumptions. When extended to real networks of varying size and connectivity, the IHC remained robust, with performance closely tied to whether the network lay below, near, or above criticality.

Despite these strengths, several limitations remain. The incentive parameter $\beta$ was treated generically as a proxy for incentive strength. While this captures the broad effect of increasing rewards on recommendation probabilities, the precise relationship between budget allocation, incentive mechanisms, and agent behavior remains an open problem \cite{abdelazeem2022effectiveness,zheng2011task}. Likewise, while we modeled heterogeneity in skills and job requirements, richer labor-market data would enable more realistic calibration of application and hiring probabilities. Finally, although we explored classical random, scale-free, and homophilic networks, other structural features, such as clustering, community structure, and temporal dynamics, may substantially influence outcomes and warrant further investigation.

Future work should therefore focus on three directions: i) grounding parameters in large-scale recruitment or recommendation datasets to refine calibration and use state-of-the-art calibration techniques \cite{yakovleva2019predict}; ii) extending the incentive component to model budget-sharing and mechanism design explicitly; and iii) testing the IHC on broader classes of empirical networks. Beyond recruitment, the IHC could also be applied to other coordinated task completion problems, such as collective mobilization, marketing campaigns, or information-seeking tasks, where incentivized diffusion plays a central role. Although the IHC is motivated by practical application, it provides a micro-level building block that complements broader data-driven agent-based approaches to labor dynamics \cite{pangallo2024data}.

The IHC model makes both theoretical and practical contributions at the intersection of network science and coordinated task completion, utilizing economic incentives. It enriches classical diffusion models with halting behavior, captures a wide range of recruitment-relevant regimes, and shows empirical validity against benchmark experiments. By bridging theoretical modeling with real-world applicability, the IHC framework provides a foundation for more effective and economically grounded strategies for engaging passive participants in recruitment and beyond.

% \section*{Abbreviations}
% \begin{itemize}
%     \item \textbf{IHC model}: Independent halting cascade model
%     \item \textbf{IC model}: Independent cascade model
%     \item \textbf{ER network}: Erdos-Renyi (random) network
%     \item \textbf{BA network}: Barabasi-Albert (random) network
% \end{itemize}

\section*{Declarations}

\subsection*{Ethics approval and consent to participate}
Not applicable

\subsection*{Consent for publication}
The authors provide their full consent for the publication of this manuscript in the Journal of Complex Networks.

\subsection*{Availability of data and material}
All the (Python) code and reproducible results are available in our GitHub repository: \url{https://github.com/blas-ko/IndependentHaltingCascadeModel}. Empirical network data are publicly available in the Netzschleuder network repository\footnote{\url{https://networks.skewed.de/}}. In particular, we used networks ``\texttt{uni\_email}'', ``\texttt{copenhagen-sms}'', and ``\texttt{Twitter\_15m}''.

\subsection*{Competing interests}
The authors declare no competing interests.

\subsection*{Funding}
This paper has been carried out within the framework of the Recovery, Transformation and Resilience Plan funds, financed by the European Union (Next Generation). Through the grant «Cátedras ENIA 2022 para la creación de cátedras universidad-empresa en IA» - AImpulsa: Cátedra UC3M-Universia de Economía del Dato y la Inteligencia Artificial Responsable aplicada a la Creación Exponencial de Valor. Manuel Cebrian acknowledges additional support from projects PID2022-137243OB-I00 and PID2023-150271NB-C21, financed by MCIN/AEI/10.13039/501100011033 and ERDF ``A way of making Europe,'' as well as project TSI100922-2023-0001 under the Convocatoria Cátedras ENIA 2022. He also appreciates the support from the ``Convocatoria de la Universidad Carlos III de Madrid de Ayudas para la recualificación del sistema universitario español para 2021-2023.'' 

\subsection*{Authors' contributions}
All authors conceptualized this research. BK came up with the idea of the IHC model, implemented all the code, and performed all the data analysis and simulations. BK and MC wrote the main bulk of this manuscript, while IU and RL helped with the editing and feedback of the first version of this manuscript. RL supervised the project.

\subsection*{Acknowledgments}
The authors would like to thank Maria Carvajal and Catenon TalentHackers for fruitful discussions. We acknowledge OpenAI's ChatGPT and Anthropic's Claude, which provided editorial assistance with this manuscript. 

%%%%%%%%%%%%%% REFERENCES %%%%%%%%%%%%%%%%
% \bibliographystyle{plain}
% \bibliographystyle{unsrt} 
\bibliographystyle{IEEEtran}
\bibliography{references}

% %%%%%%%%%%%%%% APPENDICES %%%%%%%%%%%%%%%%
\begin{appendices}
    %%% APPENDICES %%%
\section{Analytic success rate of the direct recommendation system}
\label{app:direct_success}

In this Appendix, we derive the probability of a successful hiring event within the direct recommendation system. This system is characterized by a single recommendation step initiated by a central spreader, connected to a fraction $\rho$ of the total population of size $N$. Consequently, its dynamics are independent of the underlying network topology and rely exclusively on the recommendation, application, and hiring probabilities.

As discussed in Section \ref{sec:direct_recs}, direct recommendation system depends on five key parameters: $N$ (population size), $\rho$ (fraction of the population reached by the central recommender), $p_r$ (recommendation probability), $n_\nu$ (number of requirements for a vacancy), and $\mu_s$ (average number of skills per agent, Poisson parameter for skillsets).

To compute the probability of a successful hire, we consider three components: (i) the probability that an agent is hirable, (ii) the probability that hirable agents are reachable by the direct recommender, and (iii) the probability that at least one of these agents is successfully recommended. A hirable agent is one who meets all requirements of a given job vacancy (see Section \ref{sec:skills_vacancies}).
Thus, we may write:
\begin{align}
    \mathbb{P}(\text{success}; N, \rho, p_r, \mu_s, n_\nu) 
    &= \sum_{L=0}^N P_{\text{hir}}(L; N,\mu_s,n_\nu) 
    \sum_{l=0}^L P_{\text{reach}}(l,L; N,\rho)\, P_{\text{success}}(l; p_r),
    \label{eq:direct_analytical_solution}
\end{align}
where $P_{\text{hir}}$ is the probability of having $L$ hirable agents, $P_{\text{reach}}$ is the probability that $l$ of them are reachable by the recommender, and $P_{\text{success}}$ is the probability that at least one is successfully recommended.

The last term is
\begin{equation}
    P_{\text{success}}(l;p_r) = 1 - (1 - p_r)^l,
    \label{eq:successful_hire}
\end{equation}
which is the probability of at least one success among $l$ trials with probability $p_r$.

The middle term is
\begin{equation}
    P_{\text{reach}}(l,L; N,\rho) = \frac{ \binom{L}{l} \binom{N - L}{ \rho N - l} }{ \binom{N}{\rho N} },
\end{equation}
which is the probability that exactly $l$ of the $L$ hirable agents fall within the $\rho N$ agents reached by the recommender.

Finally, the first term is
\begin{equation}
    P_{\text{hir}}(L; N,\mu_s,n_\nu) = \binom{N}{L} p_{\mu_s}^L (1 - p_{\mu_s})^{N-L},
    \label{eq:hirable_sampling}
\end{equation}
where $p_{\mu_s}$ is the probability that a random agent is hirable, depending on $\mu_s$, $n_\nu$, and $N$.

Determining $p_{\mu_s}$ requires modeling skills and requirements (see Section \ref{sec:skills_vacancies}).
For a universe of $K$ skills, the probability that an agent $u$ with $|s_u| \sim \text{Poisson}(\mu_s)$ satisfies a vacancy of size $n_\nu$ is:
\begin{equation*}
    \mathbb{P}( \nu \subseteq s_u | |\mathcal{S}| = K) = \sum_{k=0}^K p_{\text{Poisson}}(k;\mu_s)\, \frac{\binom{k}{n_\nu}}{\binom{K}{n_\nu}}.
\end{equation*}
Since $K$ is not fixed, we approximate it by the maximum skillset size in the population, $K = \max\{ |s_u|, n_\nu\}$, with distribution determined by extreme-value theory applied to $|s_u| \sim \text{Poisson}(\mu_s)$. Substituting yields:
\begin{align}
    p_{\mu_s}(\mu_s, n_\nu, N) &= \sum_{K=0}^\infty \Big( F_{\text{Poi}}(K;\mu_s)^N - F_{\text{Poi}}(K-1;\mu_s)^N \Big) 
    \left( \sum_{k=0}^K p_{\text{Poi}}(k;\mu_s) \frac{\binom{k}{n_\nu}}{\binom{K}{n_\nu}} \right).
\end{align}

Therefore, the complete expression for the success rate of the direct recommendation system is
\begin{equation}
    \mathbb{P}(\text{success}; N, \rho, p_r, \mu_s, n_\nu) 
    = \sum_{L=0}^N 
    \underbrace{\binom{N}{L} p_{\mu_s}^{L} (1 - p_{\mu_s})^{N-L}}_{\text{prob. of $L$ hirable agents}} 
    \sum_{l=0}^L 
    \underbrace{\frac{ \binom{L}{l} \binom{N - L}{ \rho N - l} }{ \binom{N}{\rho N} }}_{\text{$l$ reachable by recommender}} 
    \; \underbrace{\bigl( 1 - (1 - p_r)^l \bigr)}_{\text{at least one is activated}} .
    \label{eq:direct_hire_prob}
\end{equation}

% \begin{equation}
%     \mathbb{P}(\text{successful hire}; N, \rho, p_r, \mu_s, n_\nu) 
%     = \sum_{L=0}^N \binom{N}{L} p_{\mu_s}^L (1 - p_{\mu_s})^{N-L} 
%     \sum_{l=0}^L \frac{ \binom{L}{l} \binom{N - L}{ \rho N - l} }{ \binom{N}{\rho N} } \bigl( 1 - (1 - p_r)^l \bigr).
%     \label{eq:direct_hire_prob}
% \end{equation}

\subsection{Computation approximations}

Equation \ref{eq:direct_hire_prob} is exact but involves computationally expensive sums. Specifically, the double sum over $L$ and $l$ scales as $\mathcal{O}(N^2)$, and $p_{\mu_s}$ requires an infinite sum over $K$. In practice, both can be truncated with negligible error. For $P_{\text{hir}}$, a Binomial distribution with mean $\mu_{hir} = N p_{\mu_s}$ and standard deviation $\sigma_{hir} = \sqrt{N p_{\mu_s}(1-p_{\mu_s})}$, we restrict $L$ to the interval $[\mu_{hir} - 2.5\sigma_{hir}, \mu_{hir} + 2.5\sigma_{hir}]$, which covers about $98\%$ of the mass. Similarly, for $p_{\mu_s}$, we truncate $K$ at the smallest value $K_{max}$ such that $\sum_{K=0}^{K_{max}} \mathbb{P}(\max_u |s_u| = K) \geq 0.98$. These approximations reduce computation to a tractable number of terms while retaining high accuracy.

\section{Characterization of the homogeneous IHC dynamical regimes}\label{app:ihc_analytical}

To complement the simulations, we provide a branching-process approximation of the IHC model under homogeneous parameters \cite{athreya2012branching}. In this setting, all agents share the same recommendation, application, and hiring probabilities $p_r$, $p_a$, and $p_h$. We assume that the population size $N$ is large, that most agents remain passive, and that each agent has a finite expected degree $\langle k \rangle$. Under these conditions, the cascade can be approximated as a branching process where each activated agent produces a random number of children according to the Binomial distribution.

\paragraph{Model setup.}
A root agent does not apply but instead attempts $\langle k \rangle$ independent recommendations, each succeeding with probability $p_r$. The number of recommended neighbors it activates is therefore distributed as
\[
M \sim \mathrm{Bin}(\langle k \rangle, p_r).
\]
Each activated neighbor then behaves independently:
\begin{enumerate}
    \item with probability $p_a$, it applies, halting with probability $p_h$ (otherwise failing without producing children);
    \item with probability $1-p_a$, it does not apply and instead performs $\langle k \rangle$ new recommendation trials, each succeeding with probability $p_r$.
\end{enumerate}
This recursive structure defines the offspring distribution that governs cascade dynamics in the homogeneous IHC.

\begin{proposition}[Direct-halting boundary]\label{prop:direct_halt}
An activated neighbor halts immediately with probability $q = p_r p_a p_h$. Since the number of halts in the first generation is $H \sim \mathrm{Bin}(\langle k \rangle, q)$, the expected number of immediate halts is
\begin{equation*}
    \mathbb{E}[H] = \langle k \rangle p_r p_a p_h.
\end{equation*}
The \emph{direct-halting boundary} is defined by
\begin{equation*}
    \langle k \rangle p_r p_a p_h = 1,
\end{equation*}
separating parameter regions where halts are expected in the first generation from those where cascades must propagate further to succeed.
\end{proposition}

\begin{proposition}[Failure boundary]\label{prop:failure}
Let $U \in [0,1]$ be the probability that a single activated agent produces no halt at any depth. To compute $U$, we consider the agent’s first decision:
\begin{equation*}
    U = \underbrace{p_a(1-p_h)}_{\text{applies but fails}}
        + \underbrace{(1-p_a)\,\bigl(1-p_r + p_r U\bigr)^{\langle k \rangle}}_{\text{does not apply, all children fail}}.
\end{equation*}
The first term corresponds to applying unsuccessfully, while the second corresponds to not applying, and every child subtree also failing.  

The root creates $M_0 \sim \mathrm{Bin}(\langle k \rangle, p_r)$ children, and overall failure occurs if all of their subtrees fail:
\begin{equation*}
    P_{\text{fail}}(p_a,p_r) = \bigl(1-p_r + p_r U\bigr)^{\langle k \rangle}.
\end{equation*}
We define the \emph{failure boundary} as the set of parameter pairs $(p_a,p_r)$ for which
\begin{equation*}
    P_{\text{fail}}(p_a,p_r) = 0.5,
\end{equation*}
so that cascades have equal probability of eventually halting successfully or dying out without success.
\end{proposition}

\section{Homophily plots}\label{app:homophily_plots} 

\begin{figure}[ht]
    \centering
    \begin{subfigure}[t]{0.45\textwidth}
        \centering
        \includegraphics[width=\textwidth]{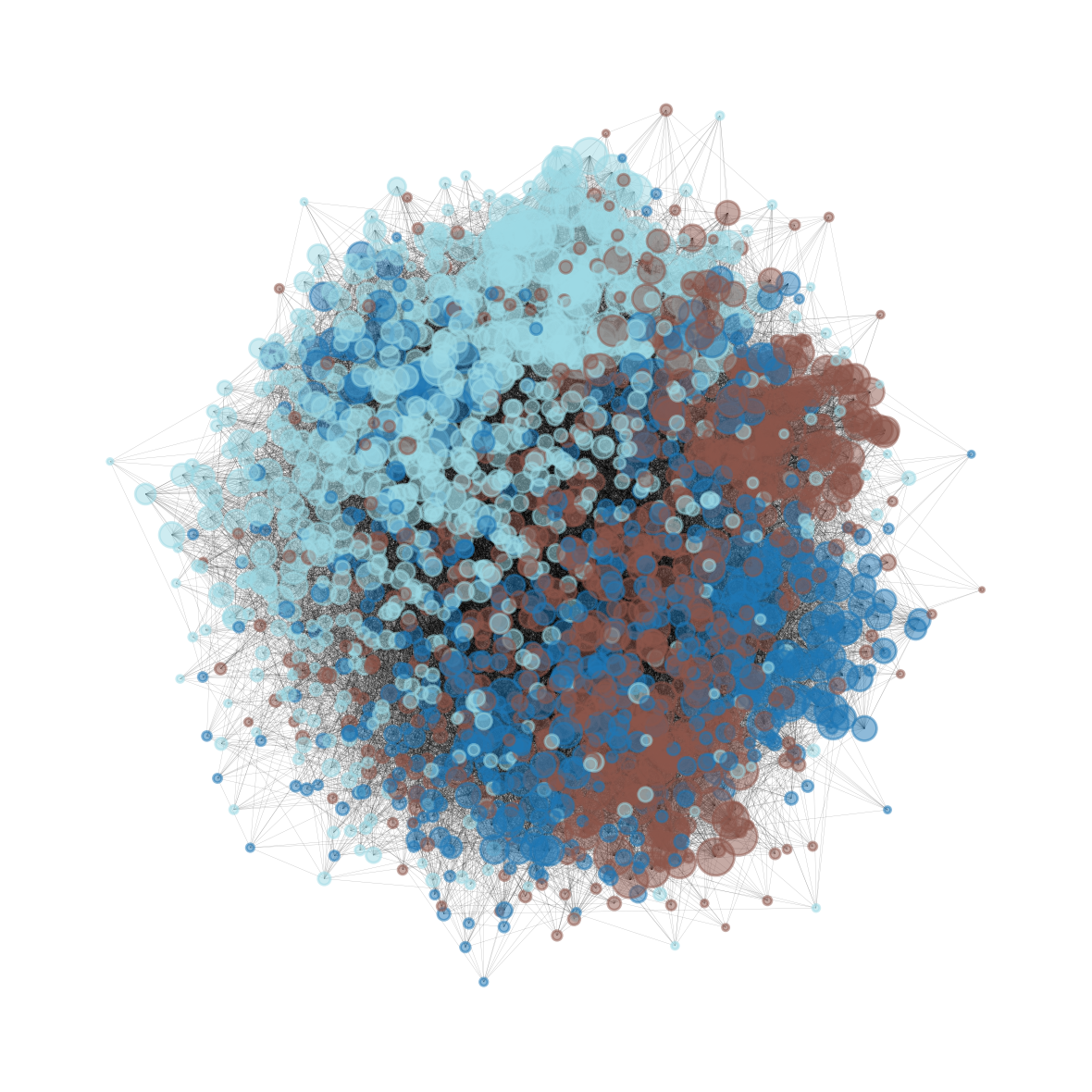}
        \caption{$\eta = 2$}
    \end{subfigure}
    \hfill
    \begin{subfigure}[t]{0.45\textwidth}
        \centering
        \includegraphics[width=\textwidth]{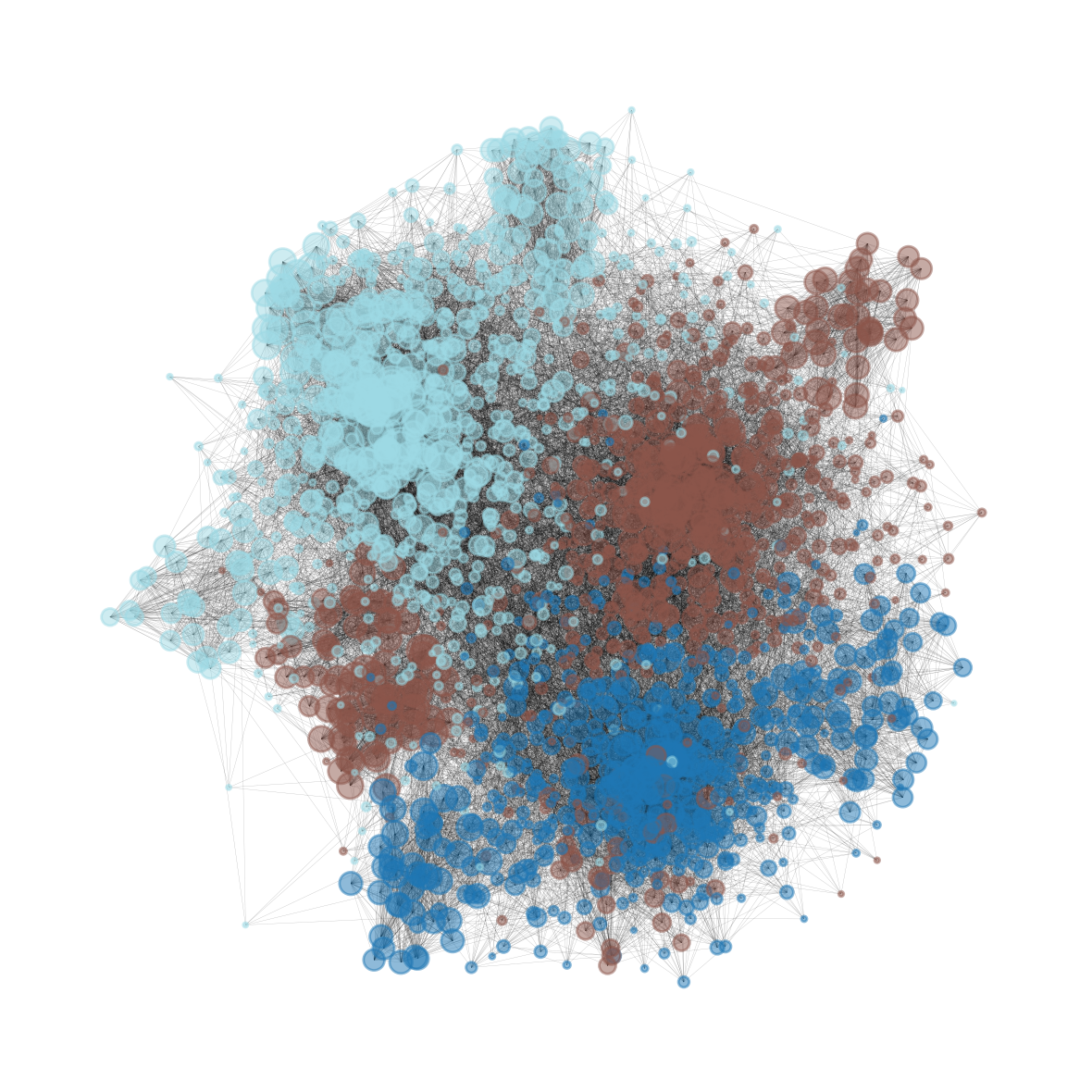}
        \caption{$\eta = 4$}
    \end{subfigure}

    \vspace{0.4cm}

    \begin{subfigure}[t]{0.45\textwidth}
        \centering
        \includegraphics[width=\textwidth]{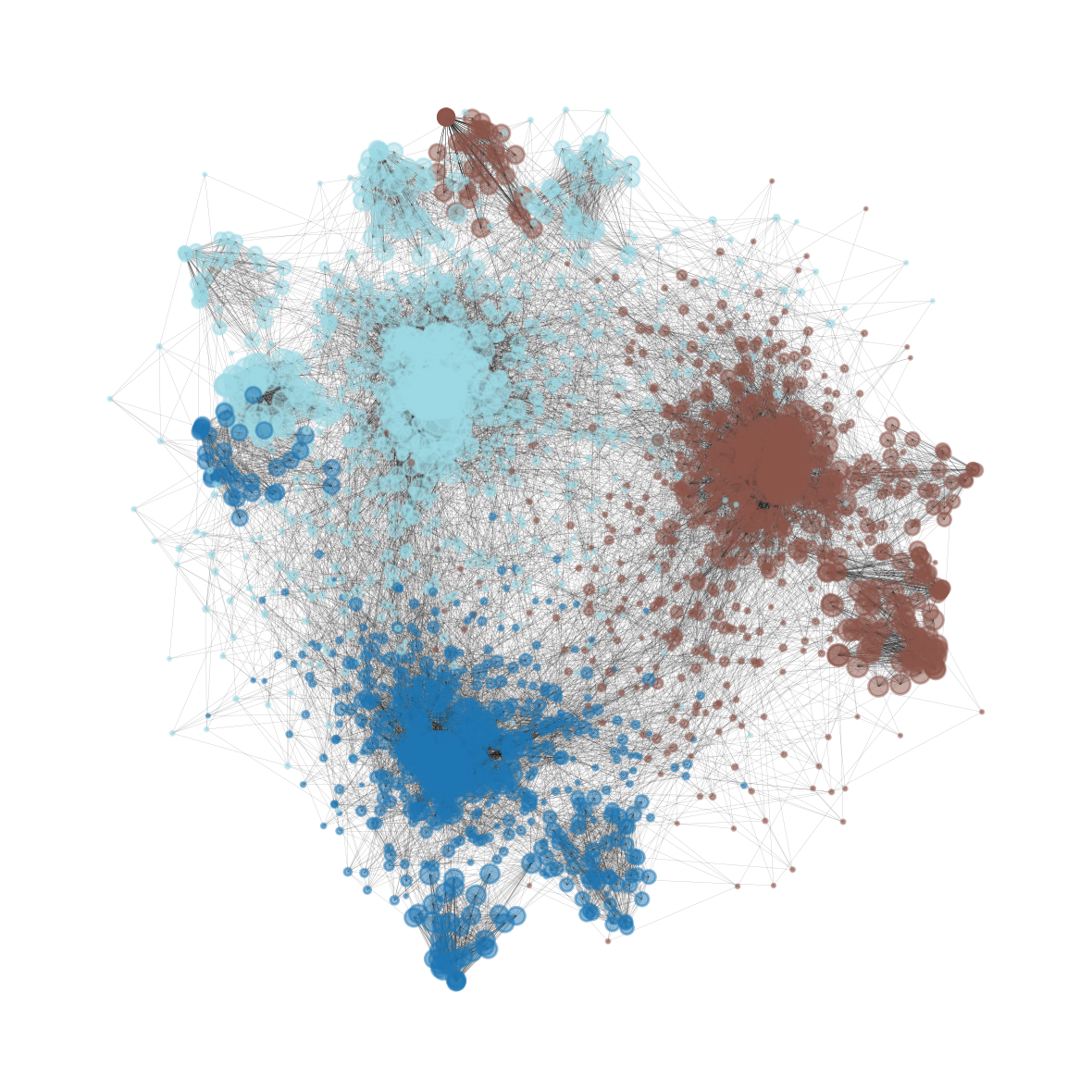}
        \caption{$\eta = 6$}
    \end{subfigure}
    \hfill
    \begin{subfigure}[t]{0.45\textwidth}
        \centering
        \includegraphics[width=\textwidth]{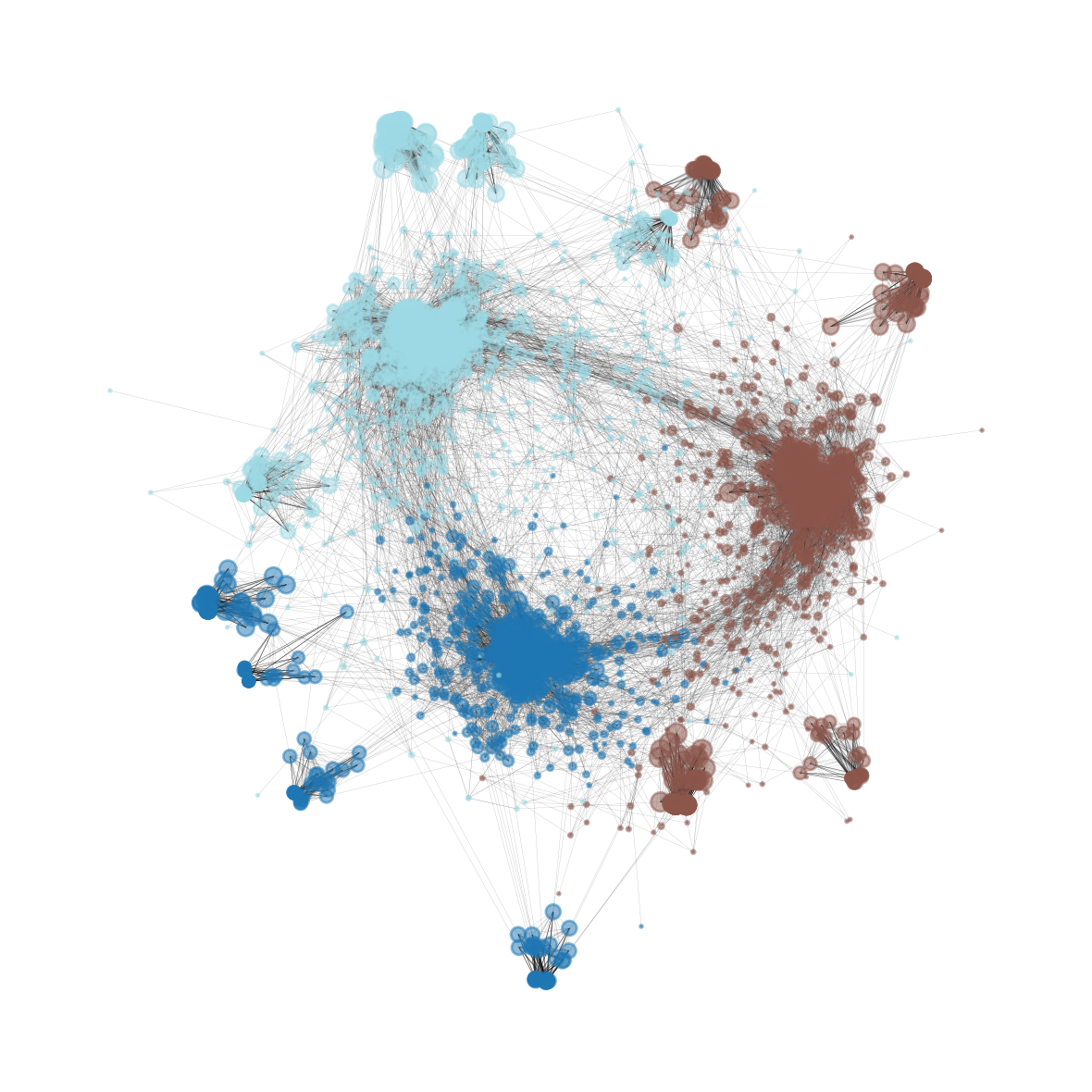}
        \caption{$\eta = 8$}
    \end{subfigure}
    \caption{Homophilic networks sampled from Eq (\ref{eq:homophily}) with $\langle k \rangle = 20$ using correlated skills ($M=3$ skill groups, Dirichlet concentration $\alpha=0.5$) for increasing levels of homophily $\eta$. Nodes are colored according to the majority group of their skill set.}
    % \\
    % \textbf{Alt text:} Panel plot showing homophilic networks sampled from increasing homophily levels, where nodes are color-coded by their skills majority group.}
    \label{fig:homophilic_networks}
\end{figure}

\begin{figure}[ht!]
    \centering
    \includegraphics[width=0.6\textwidth]{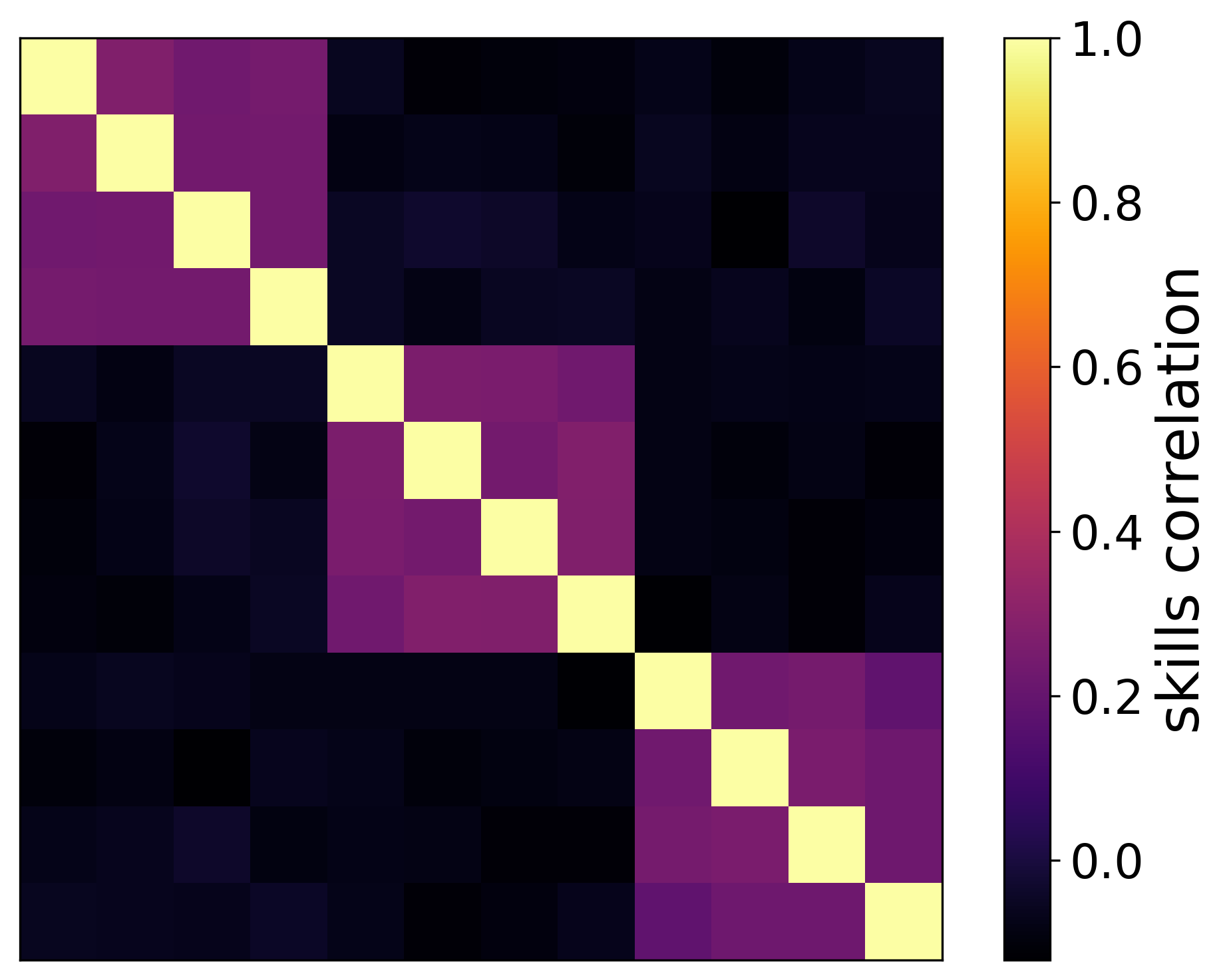}
    \caption{Correlation of agents skills after sampling correlated skills with $K = 12$ skills, $M=3$ skill groups, and concentration parameter $\alpha = 0.5$.
    % \\
    % \textbf{Alt text:} Skill correlation matrix for agents with 12 skills and 3 skill groups.
    }
    \label{fig:skill_correlations}
\end{figure}

In this Appendix, we present additional figures to strengthen the intuition behind the construction of homophilic networks based on correlated skill matching \cite{talaga2020homophily, guvenen2020multidimensional}. Figure \ref{fig:homophilic_networks} shows the resulting homophilic networks for increasing levels of homophily $\eta$, where each network has $N=2000$ nodes with average degree $\langle k \rangle = 20$. We choose $M=3$ groups of correlated skills (with concentration parameter $\alpha = 0.5$ and $12$ skills in total). Figure \ref{fig:skill_correlations} presents the correlation matrix between skills for the conditions above, highlighting the block structure of correlated skills.

% FIGURE!

% FIGURE!
\end{appendices}

\end{document}